\documentclass[epj]{svjour}
\pdfoutput=1 

\usepackage{amsmath}
\usepackage{amssymb} 
\usepackage{graphicx}

\begin{document}

\title{Contagion effects in the world network of economic activities} 

\author{V.Kandiah$^{1,2,3}$, H.Escaith$^{2,3}$  \and D.L.Shepelyansky$^{1}$}

\institute{
Laboratoire de Physique Th\'eorique du CNRS, IRSAMC, 
Universit\'e de Toulouse, UPS, F-31062 Toulouse, France
\and
World Trade Organization,
rue de Lausanne 154,
CH-1211 Gen\`eve 21,
Switzerland
\and
Opinions are personal and do not represent WTO's position
}

\titlerunning{Google matrix of the world network of economic activities}
\authorrunning{V.Kandiah, H.Escaith and D.L.Shepelyansky}

\abstract{
Using the new data from the OECD-WTO world network of economic activities
we construct the Google matrix $G$ of 
this directed network and perform its detailed analysis.
The network contains 58 countries and 37 activity sectors
for years 1995, 2000, 2005, 2008, 2009. The construction of $G$, based on 
Markov chain transitions, treats all countries on equal democratic grounds
while the contribution of activity sectors 
is proportional to their exchange monetary volume. 
The Google matrix analysis allows to obtain reliable ranking of
countries and activity sectors and to determine
the sensitivity of CheiRank-PageRank commercial balance
of countries in respect to price variations and labor cost in various countries.
We demonstrate that the developed approach takes into account 
multiplicity of network links with economy interactions between countries and activity sectors
thus being more efficient compared to the usual export-import
analysis. Our results highlight the striking increase of the influence
of German economic activity on other countries during the period 1995 to 2009
while the influence of Eurozone decreases during the same period.
We compare our results with the similar analysis of the 
world trade network from the UN COMTRADE database. 
We argue that the knowledge of network structure allows to
analyze the effects of economic influence and
contagion propagation over the world economy.
}



\date{Dated:  May  11, 2015}

\maketitle

\section{Introduction}

The matrix analysis of Input-Out transactions
had been pushed forward in the fundamental works of Leontief 
\cite{leontief1,leontief2} becoming nowadays at the heart of
modern treatment of economic relations
(see e.g. \cite{miller2009}).
The recent reports of the Organisation for Economic Co-operation and Development
(OECD) \cite{oecd2014} and of  the World Trade Organization (WTO)
\cite{wto2014} demonstrate the growing complexity of global manufactoring
activities, exchange and trade in the modern world.
Thus the advanced matrix methods are highly desirable for
the analysis of these complex systems.

In parallel, 
during the last decade the modern society generated 
enormous communication and social networks including the World Wide Web (WWW),
Wikipedia, Twitter and other directed networks (see e.g. \cite{dorogovtsev}). 
The concept of  Markov chains
provides a  powerful mathematical
approach for analysis of such networks.
In particular, the PageRank algorithm, developed by
Brin and Page in 1998 \cite{brin} for the WWW information retrieval,
became at the mathematical foundation of 
the Google search engine  (see e.g. \cite{meyer}).
This algorithm constructs the Google matrix $G$ of 
Markov chain transitions between network nodes
and allows to rank billions of web pages of the WWW.
The spectral and other properties of the Google matrix are
analyzed in \cite{arxivrmp}.
The history of the development of Google matrix methods
and their links with research in social sciences and 
works of Leontief in economy is reviewed in 
in \cite{franceschet,vignahisto}.

The results presented in \cite{wtngoogle,wtnproducts} 
for the World Trade Network (WTN), constructed from the United Nations
COMTRADE database \cite{comtrade}, show that the
Google matrix analysis is well adapted to the ranking of
world countries and trade products and to determination
of the sensitivity of trade to price variations of various
products. The new element of such an approach is an equal
(``democratic'') treatment of all countries 
independently of their richness thus being rather
different from the usual Import and Export
ranking. At the same time the contributions of various
products are considered being proportional to their trade
value (volume) contribution in the exchange flows.

\begin{figure}[!ht]
\begin{center}
\includegraphics[width=0.48\textwidth]{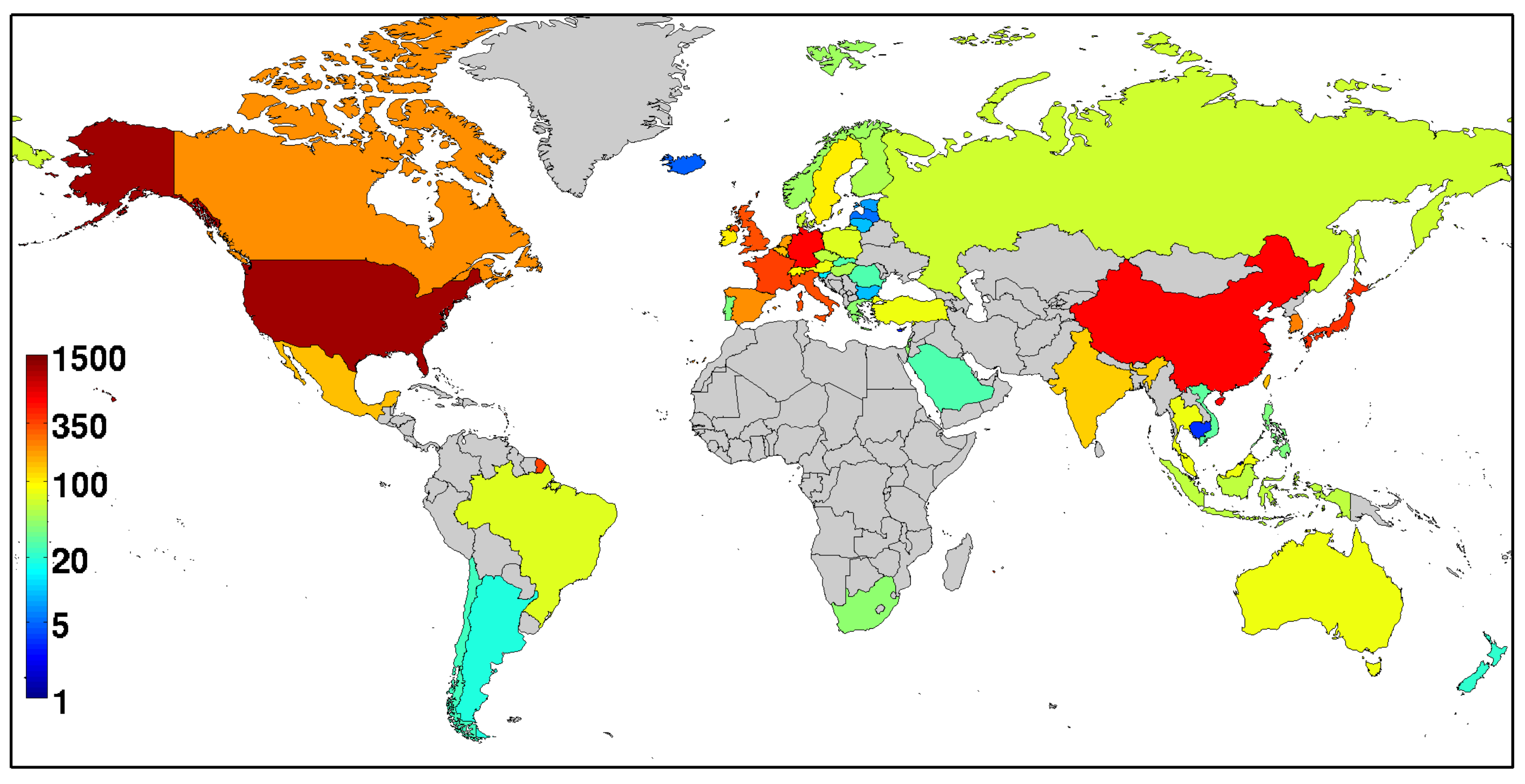}\\
\includegraphics[width=0.48\textwidth]{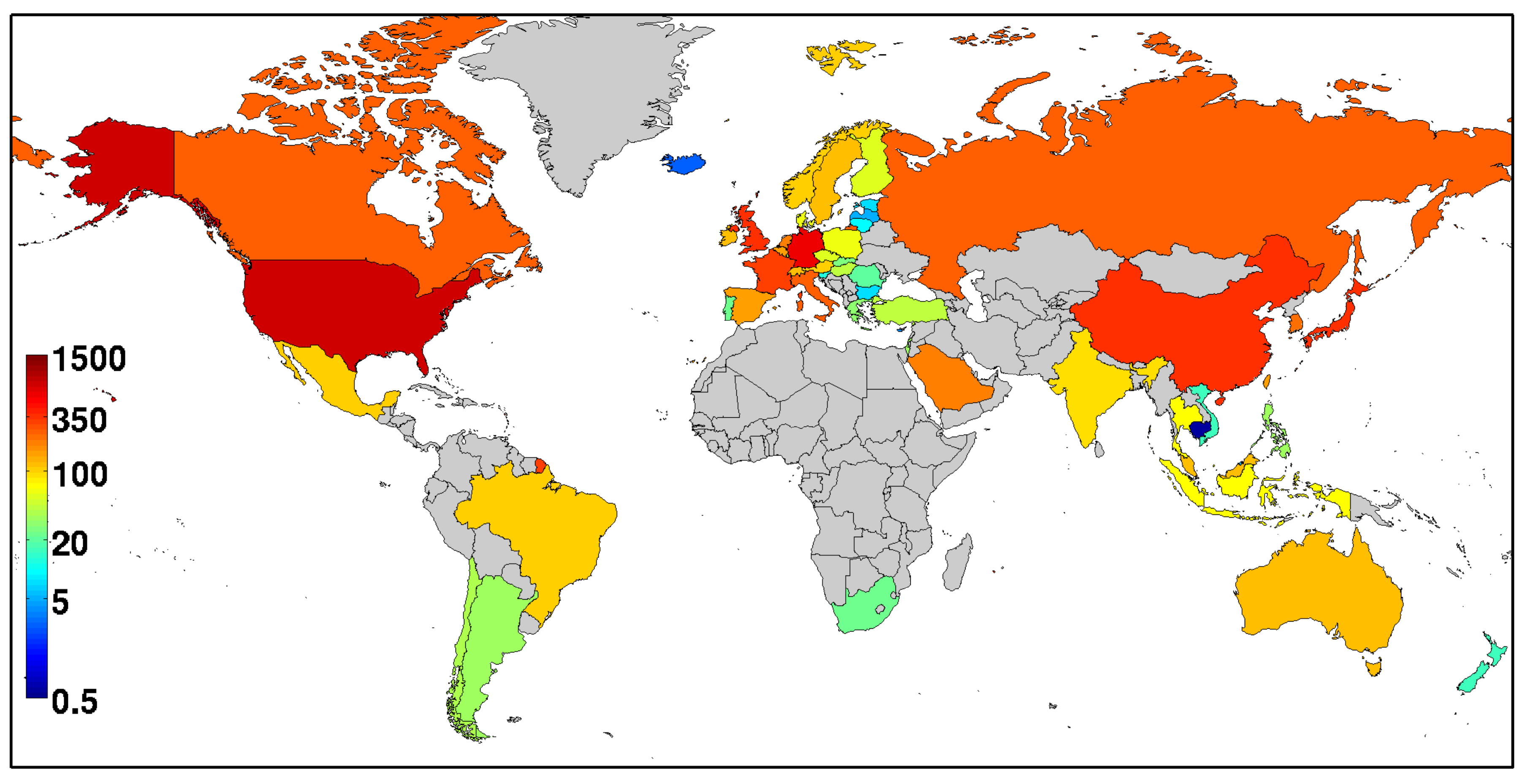}
\end{center}
\vglue -0.1cm
\caption{
World map of countries with color showing country import (top panel) 
and export (bottom panel) 
with economic activity (trade) value (volume) expressed in billions of USD 
and given by numbers at color bars;
the gray color marks countries attributed to the ROW group (rest of the world)
with exchange values $387$ (Import) and $547$ (Export) in billions of USD.
The data are shown for year 2005 with $N_c=57+1$ countries (with ROW) for the 
economic activities in all $N_s=37$ sectors. 
Country names can be found in Table~\ref{tab1} and 
in the world map of countries \cite{worldmap}.
}
\label{fig1}
\end{figure}

In this work we use the Google matrix methods for
analysis of the contagion effects on the World Network of
Economic Activities (WNEA). We use the new database of the OECD-WTO
with the network of 58 countries and 37 activity sectors.
At the difference of the World Trade matrix which report trade 
in goods between countries, the WNEA maps the imports and exports 
of intermediate goods and services between industries. 
Those globalised inter-industrial exchanges of intermediate inputs 
are one of the characteristics of the International Supply Chains 
where the production of final goods results from the combination 
of various industrial tasks that are internationally outsourced.
The first results of the Google matrix analysis 
of this database have been reported in \cite{wnea,ourwebpage}
for years 1995 and 2008 while here we analyze the time evolution
for years 1995, 2000, 2005, 2008, 2009. We show that 
our approach gives the results being different from the usual
import-export flows for individual countries
represented in Fig.~\ref{fig1} for year 2005
(world map of countries is available at \cite{worldmap}). 
The main reason of this difference is due to the fact
that the Google matrix analysis takes into account the multiplicity
of links between various nodes of the network
while the import-export approach provides only local
information without taking into account the complex 
link relations between nodes.

The new element of the OECD-WTO database is that it includes the transactions
between different activity sectors while the COMTRADE database
for multiproduct trade has no transitions between different products
(even if they exist in reality, e.g. metal
and plastic are used for production of cars).

We note that there has been a number of other investigations
of the WTN reported in 
\cite{garlaschelli,hedeem,fagiolo2,garlaschelli2010,benedictis,plosjapan,imfpaper}.
However, in this work we have the new important elements, 
developed in \cite{wtngoogle,wtnproducts,wnea}:
the analysis of PageRank and CheiRank probabilities 
corresponding to direct and inverted network flows and related to
Import and Export; democratic treatment of countries 
combined with the contributions of sectors (or products)
being proportional to their commercial exchange fractions.
We point out that the OECD-WTO TiVA database of economic activities
between  world countries and activity sectors
has been created very recently (2013) and thus this work
represents new studies of the WNEA
data evolving in time, extending the results reported
recently in \cite{wnea}.

\section{Methods and data description}

The list of $N_c=58$ countries ($57$ plus $1$ for the Rest Of the World (ROW)
is given in Table~\ref{tab1} with their flags.
Following \cite{wtngoogle} we use for countries ISO 3166-1 alpha-3 code
available at Wikipedia.
The list of sector activities with their names is given in Table~\ref{tab2} .
The sectors  are classified according to the International Standard 
Industrial Classification of All Economic Activities
(ISIC) Rev.3 \cite{isic}. 

For a given year, the TiVA data extend OECD Input/Output tables of 
economic activity expressed in terms of USD for a given year.
These data are tentative and had been released in 2013.
A next version is expected to be available in 2015.
From these data we construct
the matrix $M_{cc^\prime,ss^\prime}$
of money transfer between nodes
expressed in USD:
\begin{equation}
M_{cc^\prime,ss^\prime} = \text{transfer
from country $c^\prime$, sector $s^\prime$ to  $c,s$}
\label{eq1}
\end{equation}
Here the country indexes are $c,c^\prime=1,\ldots,N_c$ and 
activity sector indexes are $s,s^\prime=1,\ldots,N_s$
with $N_c=58$ and $N_s=37$. 
The whole matrix size is $N=N_c \times N_s = 2146$.
Here each node represents a pair of
country and activity sector, a link gives a transfer from
a sector of one country to another sector of another country.
We construct the matrix $M_{cc^\prime,ss^\prime}$
from the OECD-WTO TiVA Input/Output tables using the transposed
representation so that the volume of products or sectors flows 
in a column from line to line. In the construction of
 $M_{cc^\prime,ss^\prime}$ we exclude exchanges inside a given country
in order to highlight the trade exchange flows between countries
(elements inside country are zeros).
The method of construction of the Google matrix
from the matrix $M_{cc^\prime,ss^\prime}$ are described in \cite{wnea}
(see \cite{wtnproducts} for the COMTRADE database)
but for convenience of a reader we repeat this description here.

We  define the value of imports $V_{cs}$ and exports 
$V^{*}_{cs}$ for a given country $c$ and sector $s$  as
\begin{equation}
V_{cs}=\sum_{c^\prime,s^\prime} M_{cc^\prime,ss^\prime} \, , \,\;
V^{*}_{cs}=\sum_{c^\prime,s^\prime} M_{{{c^\prime}c},{s^\prime}s} .
\label{eq2}
\end{equation}
The import $V_c=\sum_s V_{cs}$ and export $V^*_{c} = \sum_s V^{*}_{cs}$ values 
for countries $c$  are shown on the world map of countries in Fig.~\ref{fig1}
for year 2005. We note that often one uses the notion of volume of export or import
(see. e.g. \cite{wtnproducts}) but from the economic view point it more
correct to speak about value of export or import.

In order to compare later with the PageRank and 
$\;\;\;\;\;$ CheiRank probabilities used below
we define exchange value  ranks in
the whole matrix space of dimension $N=N_c\times N_s$. Thus the
ImportRank ($\hat{P}$) and ExportRank ($\hat{P}^*$) probabilities
are given by the normalized import and export values
\begin{equation}
\hat{P}_{i} = {V_{cs}}/{V} \, , \,\;
\hat{P}^*_{i} = {V^{*}_{cs}}/{V} \, ,
\label{eq3}
\end{equation}
where $i=s+(c-1)N_s$, $i=1,\ldots,N$ and the total exchange value is
$V=\sum_{c,c^\prime,s,s^\prime} M_{cc^\prime,ss^\prime}=\sum_{c,s}V_{cs}=\sum_{cs}V^{*}_{cs}$.

The Google matrices $G$ and $G^*$ are defined as $N\times N$ real
 matrices with non-negative elements:
\begin{equation}
G_{ij}= \alpha S_{ij}+(1-\alpha) v_i e_j \, ,\; 
{G^*}_{ij}=\alpha {S^*}_{ij}+(1-\alpha) v^*_i e_j \, ,
\label{eq4}
\end{equation}
where $N=N_c\times N_s$, $\alpha \in (0,1]$ is the damping factor ($0<\alpha<1$), 
$e_j$ is the row 
vector of unit elements ($e_j=1$), and $v_i$ is a 
positive column vector called a \emph{personalization vector} 
with $\sum_i v_i=1$ \cite{meyer,wtnproducts}.
We note that the usual Google matrix corresponds to 
a personalization vector $v_i=e_i/N$ with $e_i=1$. 
In this work, following \cite{wtngoogle,wtnproducts}, we fix $\alpha=0.5$
noting that a variation of $\alpha$ in a range $(0.5,0.9)$
does not significantly affect the probability distributions of PageRank 
and CheiRank vectors \cite{meyer,arxivrmp,wtngoogle}. 
The choice of the personalization vector is specified below.
Following \cite{wtnproducts} we call this approach
the Google Personalized Vector Method (GPVM).

The matrices $S$ and $S^*$  are built from money matrices ${M}_{cc^\prime,ss^\prime}$ as
\begin{eqnarray}
 \nonumber
S_{i,i^\prime}&=&\left\{\begin{array}{cl}   
{M}_{cc^\prime,ss^\prime}/V_{c^\prime s^\prime}& 
\text{    if } V_{c^\prime s^\prime}\ne0\\ 
1/N & \text{    if } V_{c^\prime s^\prime}=0\\ 
\end{array}\right.\\
S^*_{i,i^\prime}&=&\left\{\begin{array}{cl}   
M_{{c^\prime}c,s^\prime s}/V^{*}_{c^\prime s^\prime}& 
\text{    if } V^{*}_{c^\prime s^\prime}\ne0\\ 
1/N & \text{    if } V^{*}_{c^\prime s^\prime}=0\\ 
\end{array}\right.
\label{eq5}
\end{eqnarray}
where $c,c^\prime=1,\ldots,N_c$; $s,s^\prime=1,\ldots,N_s$; 
$i=s+(c-1)N_s$; $i^\prime=s^\prime+(c^\prime-1)N_s$; 
and therefore $i,i^\prime=1,\ldots,N$. 
Here $V_{c's'}=\sum_{cs} M_{cc',ss'}$.
The sum of elements of each column of 
$S$ and $S^*$ is normalized to unity and hence the matrices $G, G^*, S, S^*$
belong to the class of Google matrices and Markov chains.
Thus $S, G$ look at the import perspective
and $S^*, G^*$ at the export side of transactions. 

PageRank and CheiRank ($P$ and $P^*$) are the right eigenvectors of
$G$ and $G^*$ matrices respectively at eigenvalue $\lambda=1$.
The equation for right eigenvectors have the form
\begin{eqnarray}
\sum_j G_{ij} \psi_j= \lambda \psi_i \, , \; 
\sum_j {G^*}_{ij} {\psi^*}_j = \lambda {\psi^*}_j \; .
\label{eq6}
\end{eqnarray}
For the eigenstate at $\lambda=1$ we use the notation
$P_i=\psi_i , P^*={\psi^*}_i$ with the normalization 
$\sum P_i = \sum_i {P^*}_i=1$. For other eigenstates we use
the normalization $\sum_i |\psi_i|^2=\sum_i |\psi^*_i|^2=1$.
The eigenvalues and eigenstates of $G, G^*$ are obtained by a direct numerical
diagonalization using the standard numerical packages.

The components of $P_i$, ${P^*}_i$ are positive. In the WWW context
they  have a meaning of probabilities
to find a random surfer on a given WWW node 
in the limit of large number of surfer jumps 
over network links \cite{meyer}.
In WNEA context nodes can be viewed and
markets with a random trader transitions between them.
We will use in the following notation of network nodes.
We define the PageRank $K$ and CheiRank $K^*$ indexes
ordering probabilities $P$ and $P^*$ 
in a decreasing order as
$P(K)\ge P(K+1)$ and $P^*(K)\ge P^*(K^*+1)$ with $K,K^*=1,\ldots,N$. 

We note that the pair of PageRank and CheiRank vectors
is very natural for economy and trade networks
corresponding to Import and Export flows.
For the directed networks the statistical 
properties of the pair of such ranking vectors
have been introduced and studied in \cite{linux,wikizzs,wtngoogle}.

We compute the reduced PageRank and CheiRank probabilities of countries 
tracing probabilities over all sectors and getting 
$P_c=\sum_{s} P_{cs}=\sum_{s}P\left(s+(c-1)N_s\right)$ 
and $P^*_c= \sum_s P^*_{cs}=\sum_{s}P^*\left(s+(c-1)N_s\right)$ 
with the corresponding $K_c$ and $K^*_c$ indexes. 
In a similar way we obtain the reduced PageRank and CheiRank probabilities
for sectors tracing over all countries and getting\\
$P_s=\sum_{c}P\left(s+(c-1)N_s\right) = \sum_{c} P_{cs}$ and \\
$P^*_s=\sum_{c}P^*\left(s+(c-1)N_s\right) = \sum_{c} P^*_{cs} $ 
with their corresponding sector indexes $K_s$ and $K^*_s$.
A similar procedure has been used for 
the multiproduct WTN data \cite{wtnproducts}. 

In summary we have $K_s,K^*_s=1,\ldots,N_s$ and 
$K_{c},K^*_{c}=1,\ldots,N_c$. A similar definition of ranks 
from import and export exchange value can be done 
in a straightforward way via probabilities
$\hat{P}_s,\hat{P}^*_s,\hat{P}_c,\hat{P}^*_c,\hat{P}_{cs},\hat{P}^*_{cs}$ and 
corresponding indexes
$\hat{K}_s,\hat{K}^*_s,\hat{K}_c,\hat{K}^*_c,\hat{K},\hat{K}^*$.

To compute the PageRank and CheiRank probabilities
from $G$ and $G^*$, keeping a ``democratic'', or equal, treatment 
of countries (independently of their richness) and at the same time
keeping the 
proportionality of activity sectors to their exchange value,
we use the Google Personalized Vector Method (GPVM) 
developed in \cite{wtnproducts} with
a personalized vector $v_i$ in (\ref{eq4}).
At the first iteration of Google matrix we  take into account 
the relative product value per country using 
the following personalization vectors for $G$ and $G^*$:
\begin{equation}
v_i = \frac{V_{cs}}{N_c \sum_{s^\prime} V_{c s^\prime}} \, , \;
v^*_i = \frac{V^{*}_{cs}}{N_c \sum_{s^\prime} V^{*}_{c s^\prime}} \, ,
\label{eq7}
\end{equation}
using the definitions (\ref{eq2}) and the relation
$i=s+(c-1)N_s$.
This personalized vector depends both on sector and country indexes.
As for the multiproduct WTN in \cite{wtnproducts}
we define the second iteration vector being proportional to the reduced 
PageRank and CheiRank vectors in sectors, obtained from the 
GPVM Google matrix of the first iteration: 
\begin{equation}
v^\prime(i) = \frac{P_s}{N_c} \, , \;
v^{\prime *}(i) = \frac{P^*_s}{N_c} \, .
\label{eq8} 
\end{equation}
In this way we keep democracy in countries but keep contribution of sectors  
proportional to their exchange value.
This second iteration personalized vectors are used in the following
computations and operations with $G$ and $G^*$
giving us the PageRank and CheiRank vectors. 
This procedure with two iterations
forms  our GPVM approach.
The difference between results obtained from
the first and second iterations is not very large, but
 the personalized vector for the second iteration
gives a reduction of fluctuations \cite{wnea}.
Below, in all Figures
we show the  GPVM results after the second iteration.

As for the WTN it is convenient to analyze the distribution of nodes on the 
PageRank-CheiRank plane $(K,K^*)$. 
In addition to two ranking indexes $K,K^*$
we use also 2DRank index $K_2$ which describes
the combined contribution of two ranks
as described in \cite{wikizzs}.
The ranking list   $K_2(i)$ is constructed by  
increasing $K \rightarrow K+1$ 
and increasing 2DRank index 
$K_2(i)$ by one if a new entry is present in the list of
first $K^*<K$ entries of CheiRank, then the one unit step is done in
$K^*$ and $K_2$ is increased by one if the new entry is
present in the list of first $K<K^*$ entries of CheiRank.
More formally, 2DRank $K_2(i)$ gives the ordering 
of the sequence of nodes, that $\;$ appear
inside  $\;$ the squares  $\;$
$\left[ 1, 1; \;K = k, K^{\ast} = k; \; \-... \right]$ when one runs
progressively from $k = 1$ to $N$.
Additionally, we analyze the distribution of nodes
for reduced indexes $(K_c,K^*_c)$, $(K_s,K^*_s)$.

\section{Results}

We apply the GPVM approach to the data sets of OECD-WTO TiVA of WNEA
and present the obtained results below.

\begin{figure}[!ht]
\begin{center}
\includegraphics[width=1\columnwidth,clip=true,trim=0 0 0 0cm]{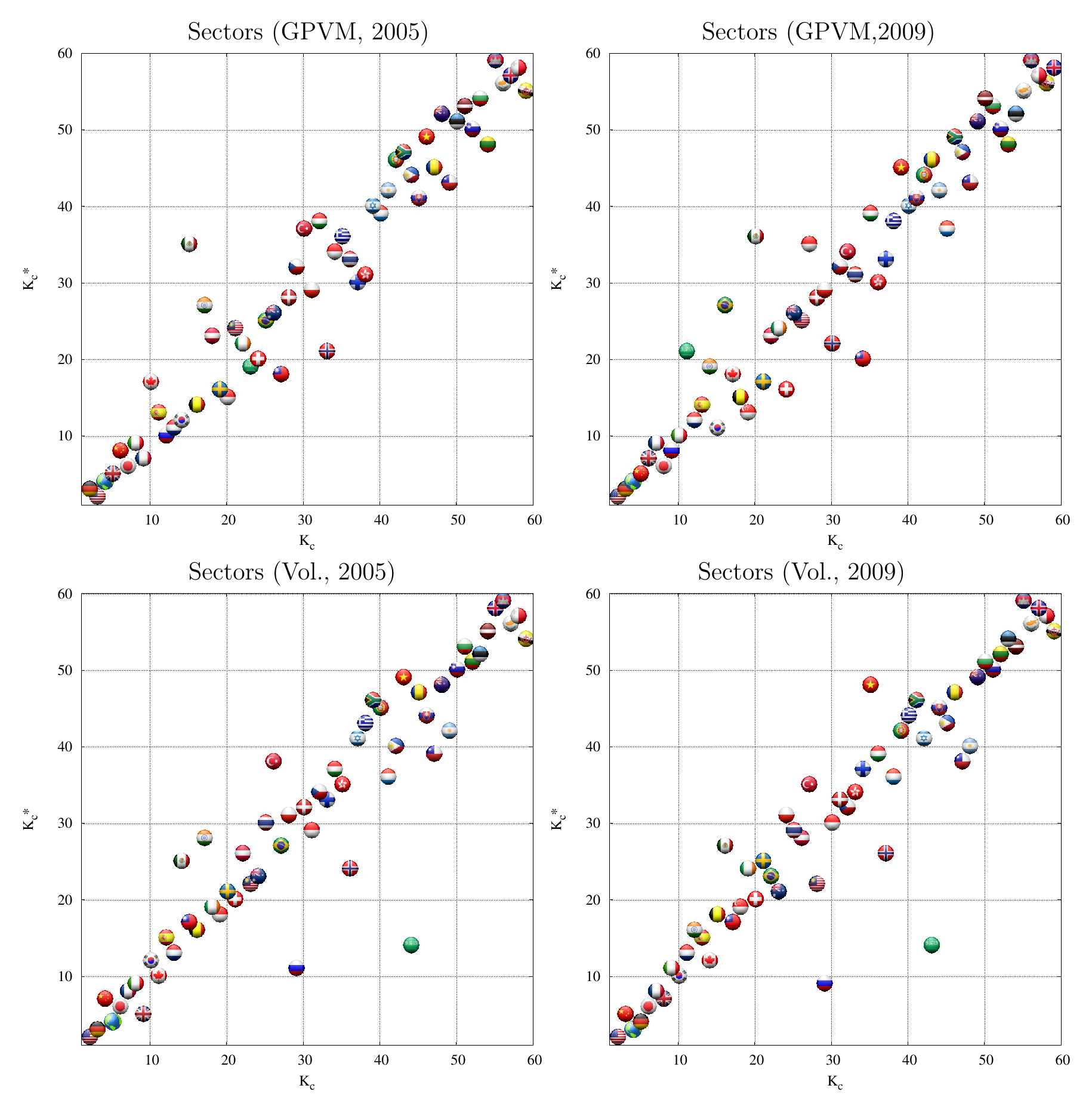}
\end{center}
\vglue -0.1cm
\caption{
Country positions on PageRank-CheiRank plane ($K_c$,$K^*_c$) 
obtained for the  WNEA by 
the GPVM analysis (top panels), ImportRank-ExportRank of 
exchange value/volume (bottom panels). 
Left (right) panels show  year 2005 (2009). 
}
\label{fig2}
\end{figure}

\subsection{Ranking of countries and sectors}

After tracing the probabilities $P(K), P^*(K^*)$ 
over sectors we obtain the distribution of world countries on the
PageRank-CheiRank plane $(K_c,K^*_c)$ presented in Fig.~\ref{fig2} 
for WNEA in years 2005, 2009. In the same figure 
we present the rank distributions 
obtained from ImportRank-ExportRank probabilities of exchange value.
For the GPVM data we see the global features already discussed in 
\cite{wtngoogle,wnea}: the countries are distributed in a vicinity of diagonal
$K_c=K^*_c$ since for each country the size of imports is correlated
with the size of exports, even if trade is never exactly balanced and some
countries can sustain significant trade surplus or deficit.
The top $20$ list of top $K_2$ countries recover $13$  of 19 countries of
$G20$  major world economies (EU is the number 20) 
thus obtaining 58\% (2005) and 63\% (2009) of the whole list.
This is close to the percentage 68\% obtained in \cite{wnea}
for year 2008. The Google ranking 
for WNEA (top panels in Fig.~\ref{fig2})
gives different  positions for specific
countries. Thus Russia and Saudi Arabia  go 
to top $K_c$ index values in PageRank
comparing to ImportRank showing that
their economies are highly sensitive to 
activity of petroleum sector. Similar 
features for these two countries 
are visible in 1995, 2008 \cite{wnea}.

\begin{figure}[!ht]
\begin{center}
\includegraphics[width=0.48\textwidth,clip=true,trim=0 0 0 0cm]{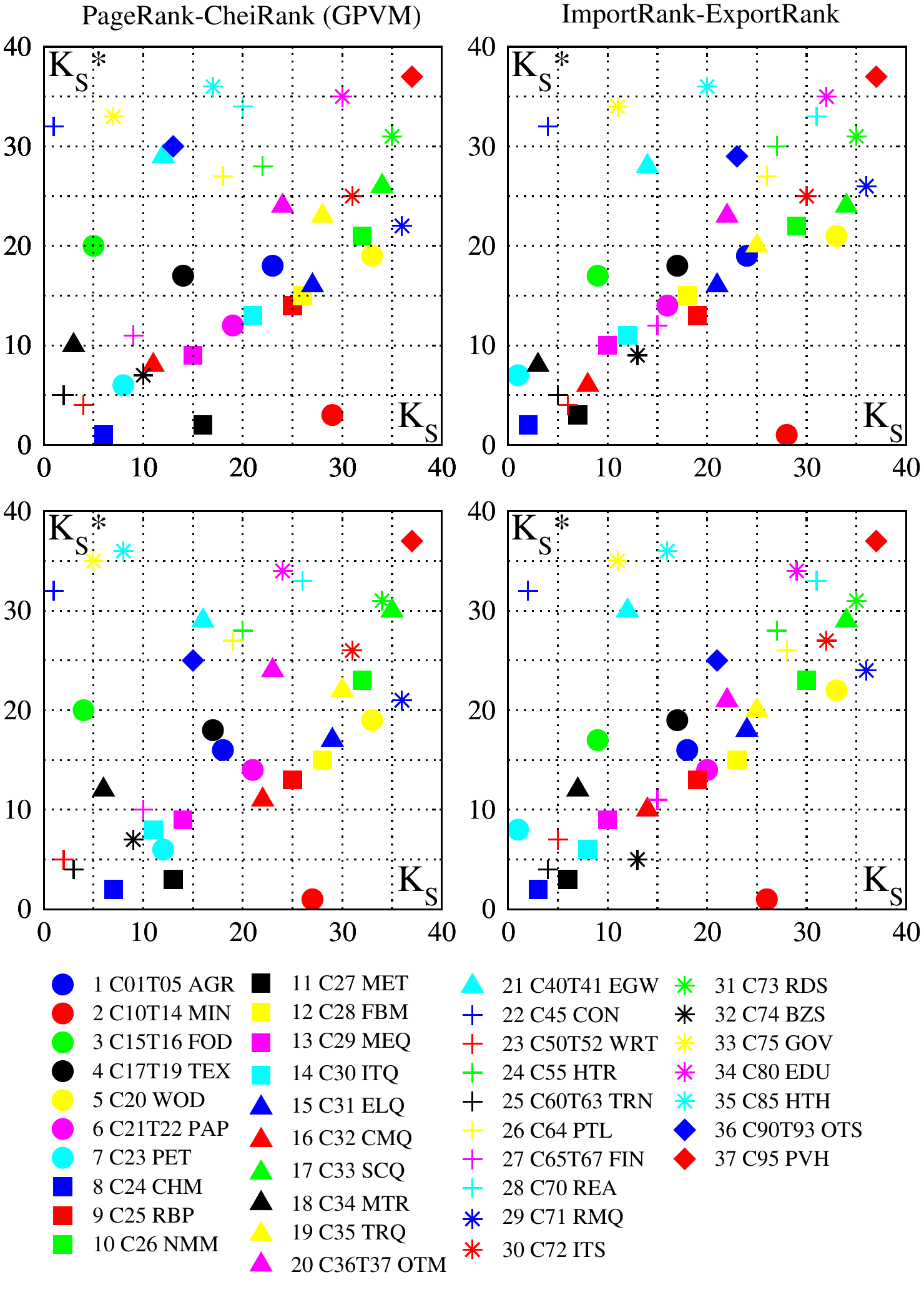}
\end{center}
\vglue -0.1cm
\caption{
Two-dimensional ranking of sectors on the ($K_s$,$K^*_s$) plane using 
the GPVM approach for PageRank and CheiRank (left panels) and 
ImportRank-ExportRank (right panels). 
Each sector is represented by its specific combination of color and symbol. 
The list of all $37$ sectors are given in Table \ref{tab2}. 
Top panels are for year 2005 and bottom panels are for year 2009.
}
\label{fig3}
\end{figure}

After tracing over countries we obtain the PageRank-CheiRank
plane $(K_s,{K_s}^*$ of activity sectors shown in Fig.~\ref{fig3}.
As in \cite{wnea} we find that some sectors are export oriented
(e.g. $s=2$ {\it C10T14 Mining} at $K_s^*=1$ in 2009)
others are import oriented 
(e.g. $s=22$ {\it  C45 CON Construction} at $K_s=1$ in 2009).
The ImportRanking gives a rather different
import leader $s=7$ {\it C23 Manufacture of coke, refined petroleum products  etc.}
with $K_s=1$ in 2009. Thus the Google ranking highlights 
highly connected network nodes while Import-Export
gives preference to high value neglecting
existing network relations between various countries and activity sectors.

\subsection{Price shocks and trade balance sensitivity}

On the basis of the obtained WNEA Google matrix we can now analyze the trade balance
in various activity sectors for all world countries.
Usually economists consider the export and import of a given country 
as it is shown in Fig.~\ref{fig1}.  Then the trade balance of a given country $c$
can be defined making summation over all sectors:  
\begin{equation}
B_c=\sum_s (P^*_{cs} - P_{cs})/\sum_s (P^*_{cs} + P_{cs}) = (P^*_{c} - P_{c})/(P^*_{c} + P_{c}) .
\label{eq9} 
\end{equation}
In economy, $P_c, P^*_c$ are defined via the probabilities of trade value
$\hat{P}_{cs}, \hat{P}^*_{cs}$  from (\ref{eq3}). In our matrix approach,
we define $P_{cs}, P^*_{cs}$ as PageRank and CheiRank probabilities. 
In contrast to the Import-Export value our approach takes into account the 
multiple network links between nodes. 

\begin{figure}[!ht] 
\begin{center} 
\includegraphics[width=1\columnwidth,clip=true,trim=0 0 0 0cm]{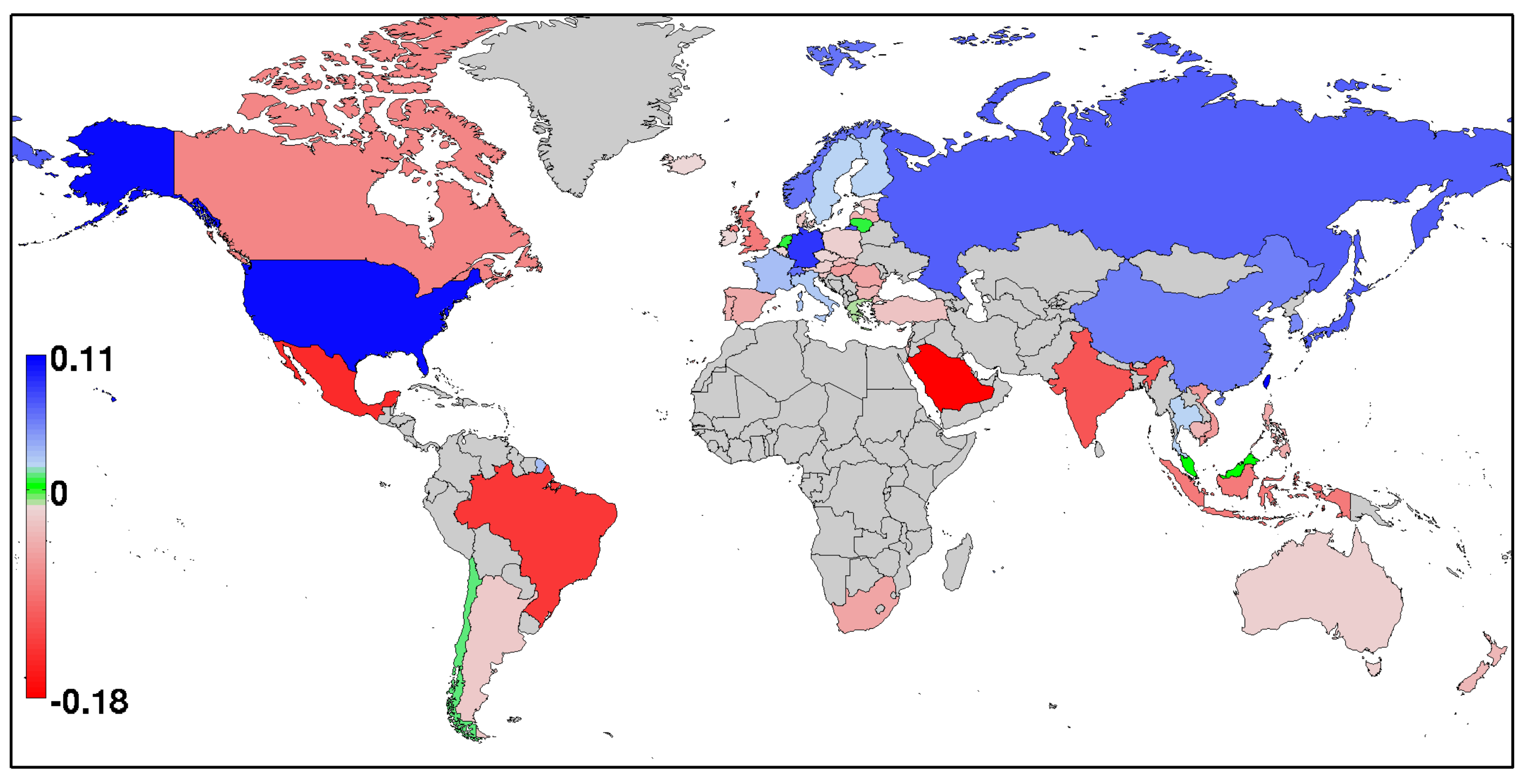} \\
\includegraphics[width=1\columnwidth,clip=true,trim=0 0 0 0cm]{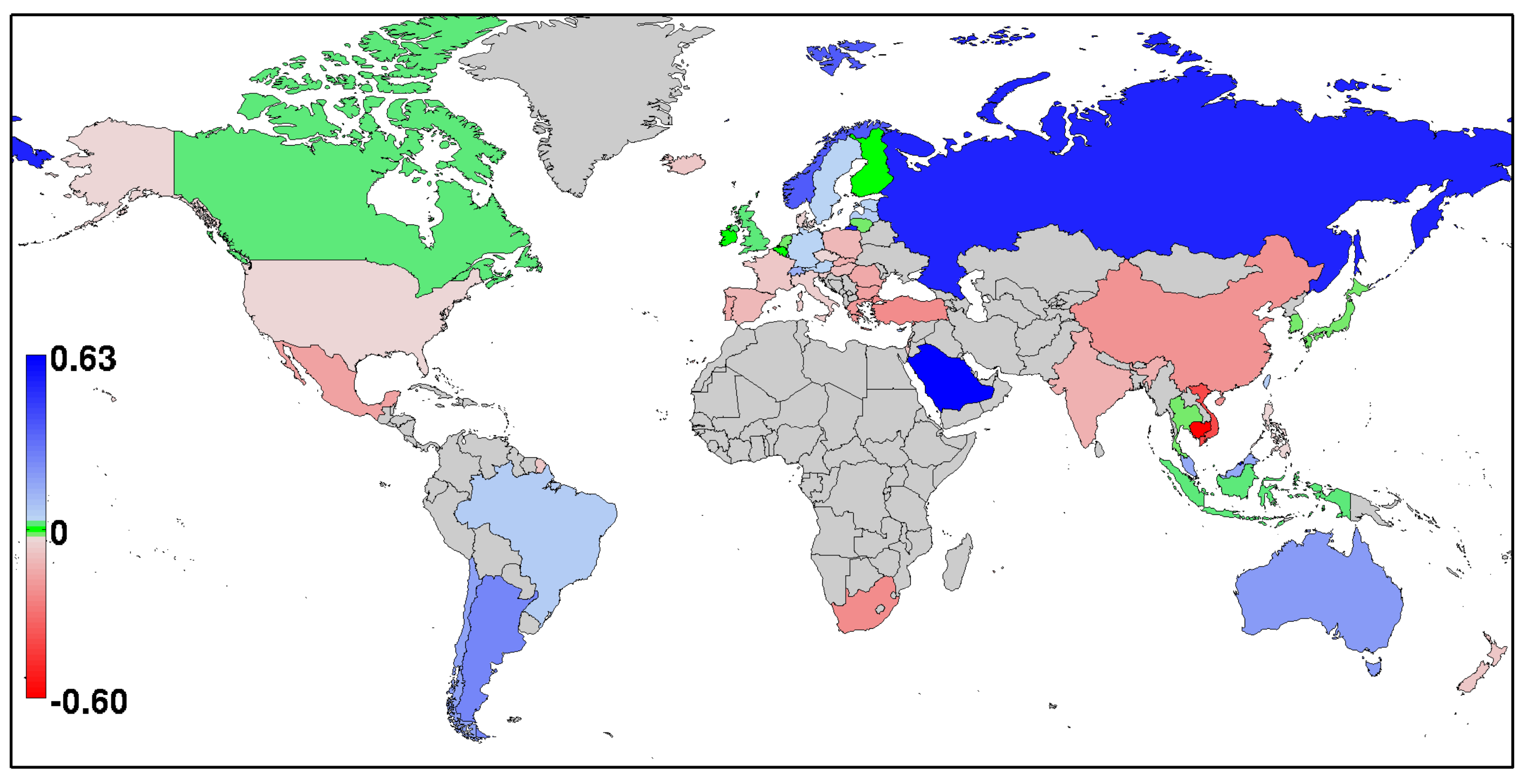}
\vglue -0.1cm
\caption {World map of CheiRank-PageRank balance $B_c=(P^*_c-P_c)/(P^*_c+P_c)$ 
determined for all ${N_c}=58$ countries in year 2009. Top panel shows 
the probabilities $P$ and $P^*$ given by PageRank and CheiRank vectors; 
the value of ROW group is $B_{c=58}=-0.0202$. Bottom panel shows 
the probabilities $P$ and $P^*$ computed from 
the  Export and Import value; the value of ROW group is $B_{c=58}=0.0637$. 
Names of the countries are given in Table \ref{tab1} and 
in the world map of countries \cite{worldmap}.}
\label{fig4}
\end{center}
\end{figure}

\begin{figure}[!ht] 
\begin{center} 
\includegraphics[width=1\columnwidth,clip=true,trim=0 0 0 0cm]{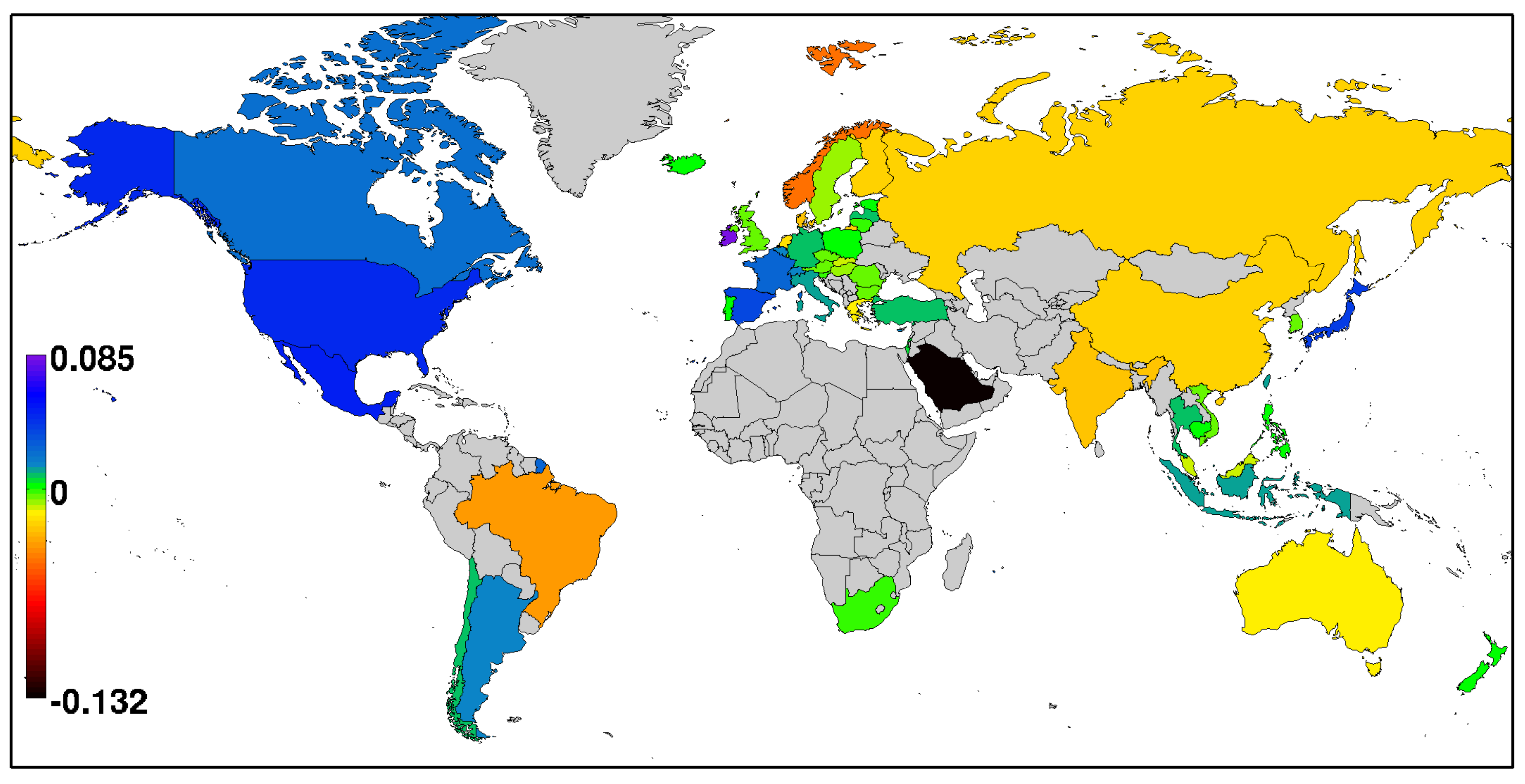}
\vglue -0.1cm
\caption {
Difference of CheiRank-PageRank balance $\Delta B_c=B_c(2009)-B_c(2008)$ 
between years 2009 and 2008 shown by color for the world countries;
for the ROW group we have $\Delta B_c = -0.043$ (gray).
Names of the countries can be found in Table \ref{tab1} and
in the world map of countries \cite{worldmap}.}
\label{fig5}
\end{center}
\end{figure}

The comparison of CheiRank-PageRank balance with
Export-Import balance for the world countries is shown in
Fig.~\ref{fig4} for year 2009. Each country is shown by color 
which is proportional to the country balance $B_{c}$ (\ref{eq9})
with the color bar given on the figure.
For Export-Import balance we see the dominance of petroleum
producing countries Saudi Arabia, Russia, Norway
with the largest values in 2009 and 2008 (see Fig.13 in \cite{wnea}).
$\;\;$ The  CheiRank-PageRank balance highlights new features
placing on the top Russia, Norway, 
Germany, China in 2008 \cite{wnea}. 
In fact in 2008, USA has now a slightly positive balance 
in top panel of Fig.13 while it was negative before
in bottom panel of same figure.
In 2009 after the world crisis there is a significant change
for CheiRank-PageRank balance in the top panel of Fig.~\ref{fig4} in 2009:
USA  takes the leading position while Saudi Arabia becomes even negative.
The variation of CheiRank-PageRank balance
$\Delta B_c=B_c(2009)-B_c(2008)$
from 2008 to 2009 is shown in Fig.~\ref{fig5}. 
The strongest positive variation is obtained by Ireland, USA and Japan,
the strongest negative variation is for Saudi Arabia
and Norway.
We see that the broad network of economic activity relations and links
makes the economies of the above countries more important in the
world economy while Saudi Arabia, with the largest positive
Export-Import balance, looses its leading position.

To analyze the sensitivity of price variation in a certain activity sector $s$
we increase from $1$ to $1+\delta_s$  the  money transfer
in the sector $s$ in $M_{cc\, ss'}$ in (\ref{eq1}), where $\delta_s$
is a dimensionless fraction variation of price in this sector.
After that the matrices $G, G^*$ are recomputed in the usual way described above
and their rank probabilities $P, P^*$ are determined.
Then we compute the derivatives of probabilities
balance $d B_c/d \delta_s$ over a price variation $\delta_s$
in a specific sector $s$. Of course, the computation is done
at small values of $\delta_s$ when the derivative
is independent of $\delta_s$ and all price variations
are kept sufficiently small.

\begin{figure}[!ht] 
\begin{center} 
\includegraphics[width=1\columnwidth,clip=true,trim=0 0 0 0cm]{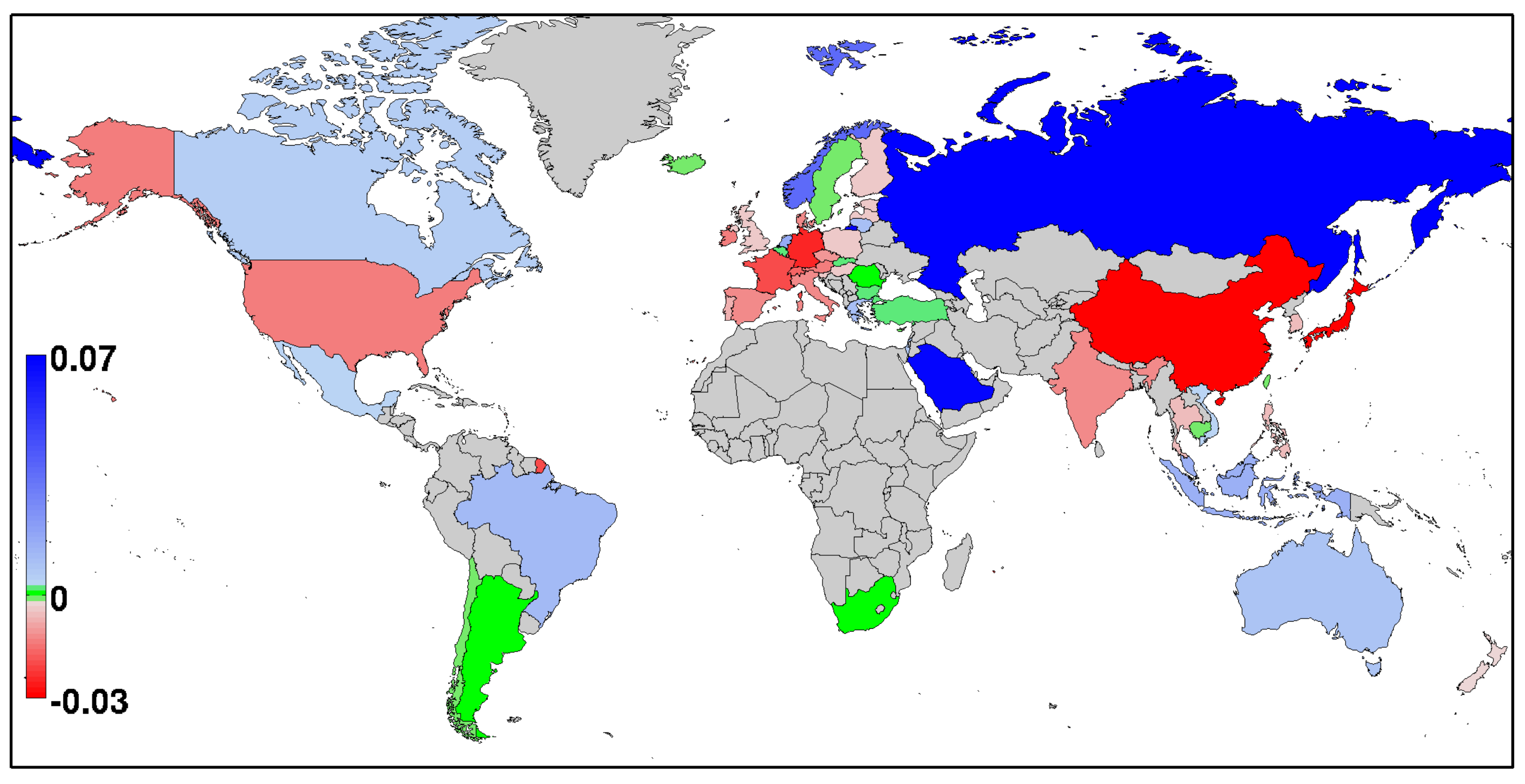} \\
\includegraphics[width=1\columnwidth,clip=true,trim=0 0 0 0cm]{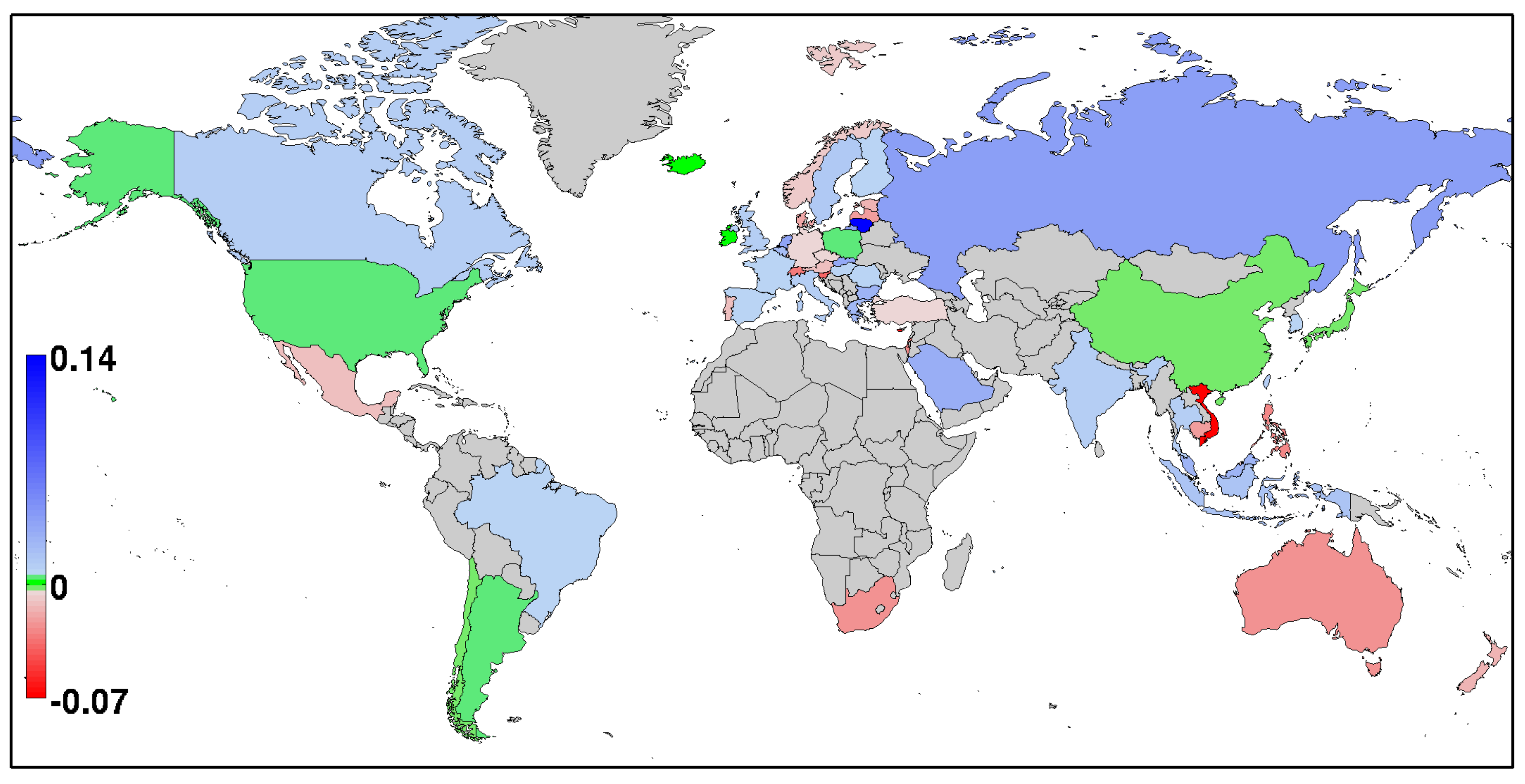}
\vglue -0.1cm
\caption {
Derivative of probabilities balance $dB_c/d\delta_7$ over price of sector
$s=7$ C23PET 
for year 2009. Top panel shows the case when $B_c$ is determined by 
CheiRank and PageRank vectors as in the top panel of Fig.\ref{fig4}; 
the value of ROW group is $dB_{58}/d\delta_7=0.0414$. Bottom panel shows the case 
when $B_c$ is computed from the Export-Import value as in the bottom panel of 
Fig.\ref{fig4}; the value of ROW group is $dB_{58}/d\delta_7=-0.0637$. 
Names of the countries can be found in Table \ref{tab1} and
in the world map of countries \cite{worldmap}.}
\label{fig6}
\end{center}
\end{figure}

\begin{figure}[!ht] 
\begin{center} 
\includegraphics[width=1\columnwidth,clip=true,trim=0 0 0 0cm]{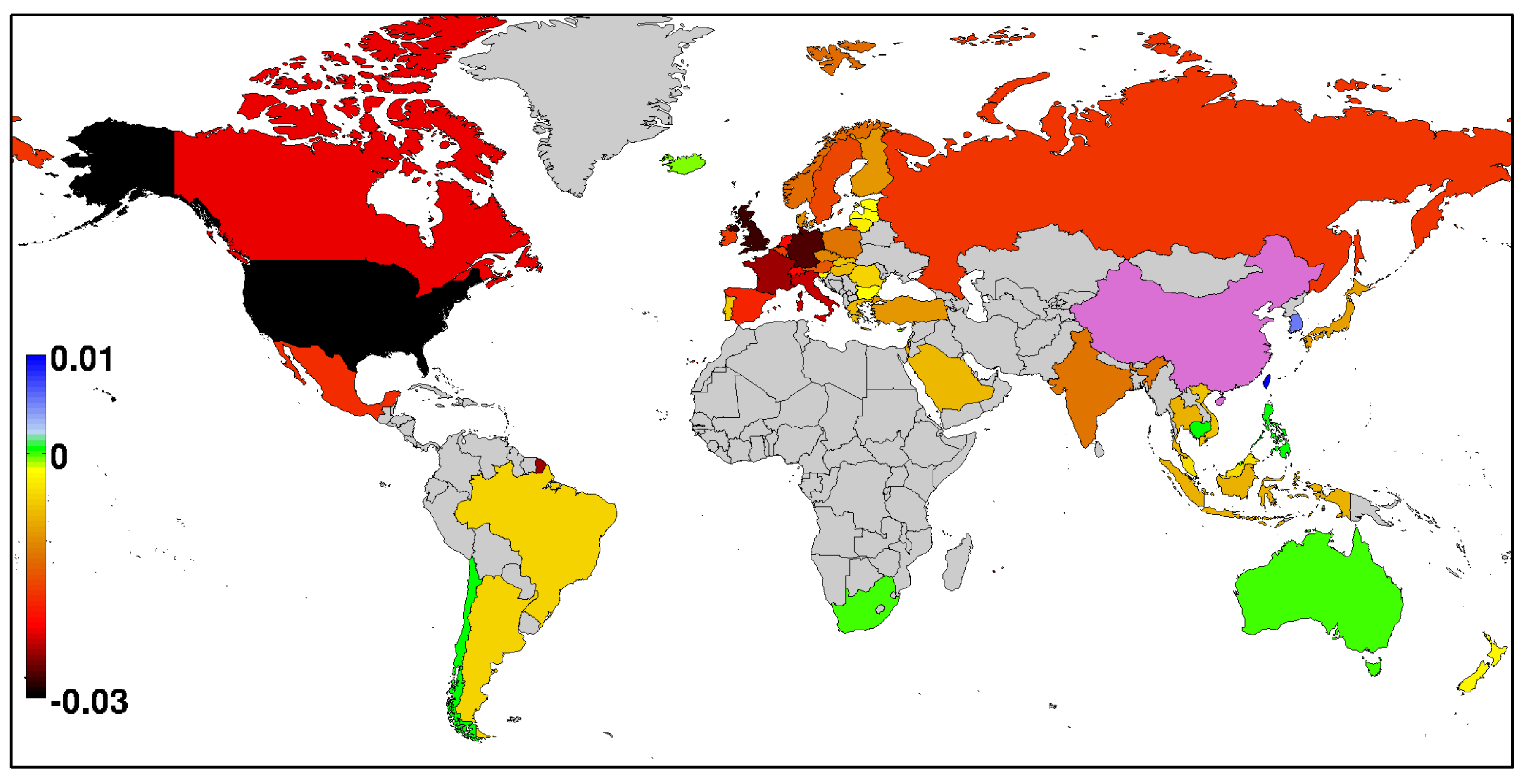} \\
\includegraphics[width=1\columnwidth,clip=true,trim=0 0 0 0cm]{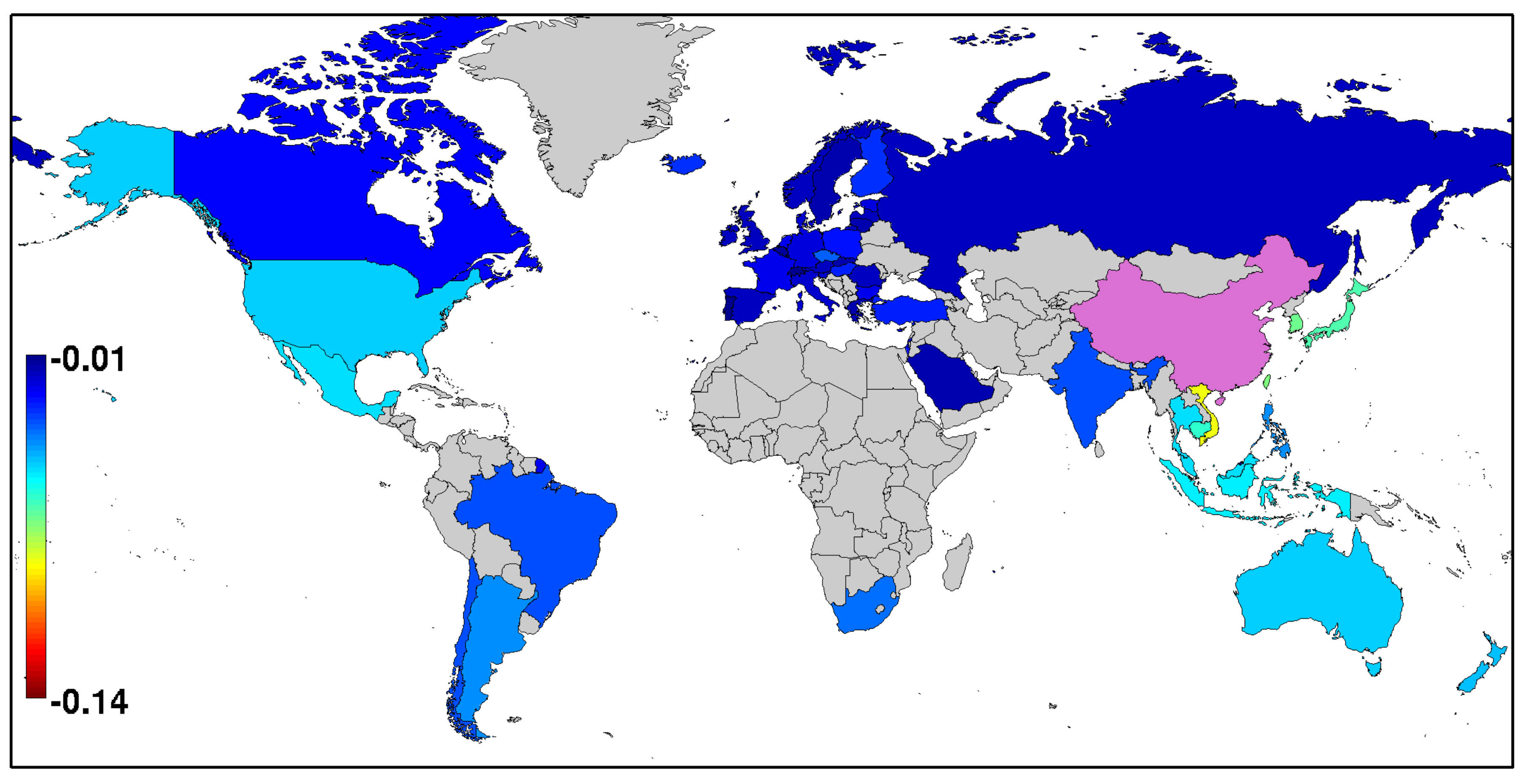}
\vglue -0.1cm
\caption {
Derivative of probabilities balance $dB_c/d\sigma_{c'}$ 
over labor cost of China $c'=37$ for year 2009. 
Top panel shows the case when $B_c$ is determined 
by CheiRank and PageRank vectors; 
here the special values are $dB_{58}/d\sigma_{37}=-0.0163$ 
for ROW group (gray) and 
$dB_{37}/d\sigma_{37}=0.3253$ for China (magenta). 
Bottom panel shows the case when $B_c$ is computed from 
the  Export-Import value; 
the special values are $dB_{58}/d\sigma_{37}=-0.0381$ for ROW group (gray) and 
$dB_{37}/d\sigma_{37}=0.4732$ for China (magenta). 
Names of the countries can be found in Table \ref{tab1} and
in the world map of countries \cite{worldmap}.}
\label{fig7}
\end{center}
\end{figure}

The sensitivity of country balance $d B_c/d \delta_7$
to price variation of sector $s=7$ 
{\it Manufacture of coke, refined petroleum products and nuclear fuel}
is shown in Fig.~\ref{fig6}. For Export-Import in bottom panel
the most sensitive
countries are  Lithuania (positive)
and Vietnam (negative).
Lithuania does not produce petroleum,
but in fact in 2008 there was a  
large oil refinery company there which had a large exportation value
(see e.g. \\
http://en.wikipedia.org/wiki/Economy\_of\_Lithuania).
The Export-Import approach shows that Russia is slightly positive,
even less positive is Saudi Arabia,
China and Germany are close to zero change, USA is only
very slightly positive.
The results of CheiRank-PageRank sensitivity (top panel)
are significantly different
showing strongly positive sensitivity for Saudi Arabia, Russia
and strongly negative sensitivity for China, Germany and Japan;
USA goes from slightly positive side in bottom panel to moderate negative one
in top panel. The CheiRank-PageRank balance 
demonstrates much higher sensitivity of Russia, Saudi Arabia
and China to price variations of $s=7$ sector
comparing to the case of Export-Import value analysis.
The economies of Germany, China and Japan are also very 
sensitive to petroleum prices that is correctly captured by our
analysis.
We consider that the CheiRank-PageRank
approach describes the economic reality 
from a new complementary angle and 
that provides new useful information about
complex trade systems.
 We also note that the highly negative sensitivity
of China to $\;\;\;$ petroleum prices has been also obtained
on the basis of Google matrix analysis of COMTRADE data 
(see Fig.21 in \cite{wtnproducts}).

It is also possible to determine the partial
balance for a given sector $s$ and given country $c$ 
and to study its sensitivity to price variations in 
a sector $s'$. We do not discuss these
characteristics here but
an interested reader can find these
results for year 2008 in \cite{wnea}.

Of course, the above derivatives over price of activity sector and labor country cost
give only an approximate consideration of effects of price variations
which is a very complex phenomenon. 
For an economic discussion of the effect of
price shocks on international production networks we address a reader to
the research performed in \cite{escaith}.
We will see below that our
approach gives  results being in a good agreement with economic realities
thus opening complementary possibilities
of economic activity analysis based on the underlying
network relations between countries and activity sectors
which are absent in the usual Import-Export consideration.
We present the results on sensitivity to sector prices and labor cost in next
subsections.

\clearpage 

\begin{figure}[!ht] 
\begin{center} 
\includegraphics[width=0.9\columnwidth,clip=true,trim=0 0 0 0cm]{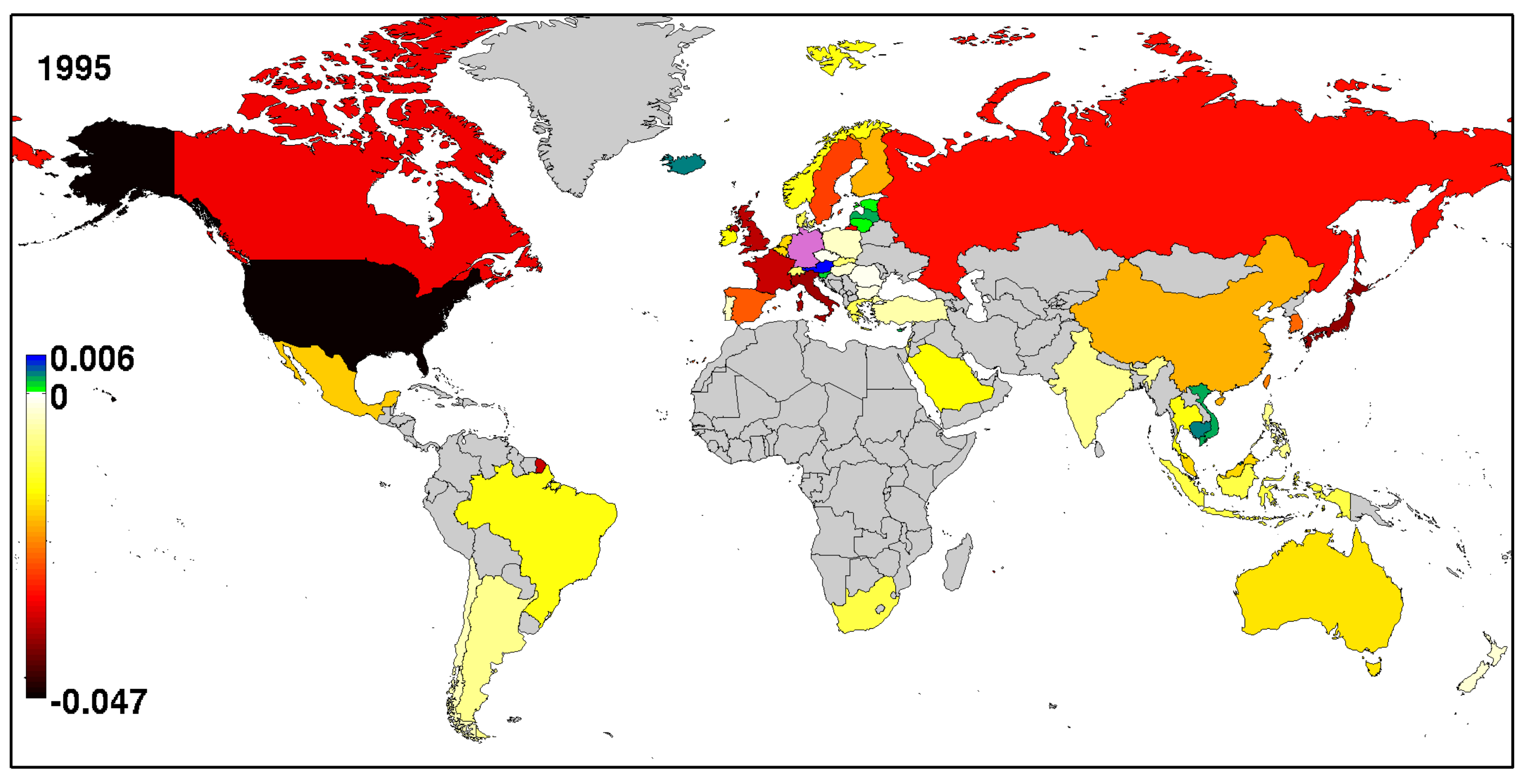} \\
\includegraphics[width=0.9\columnwidth,clip=true,trim=0 0 0 0cm]{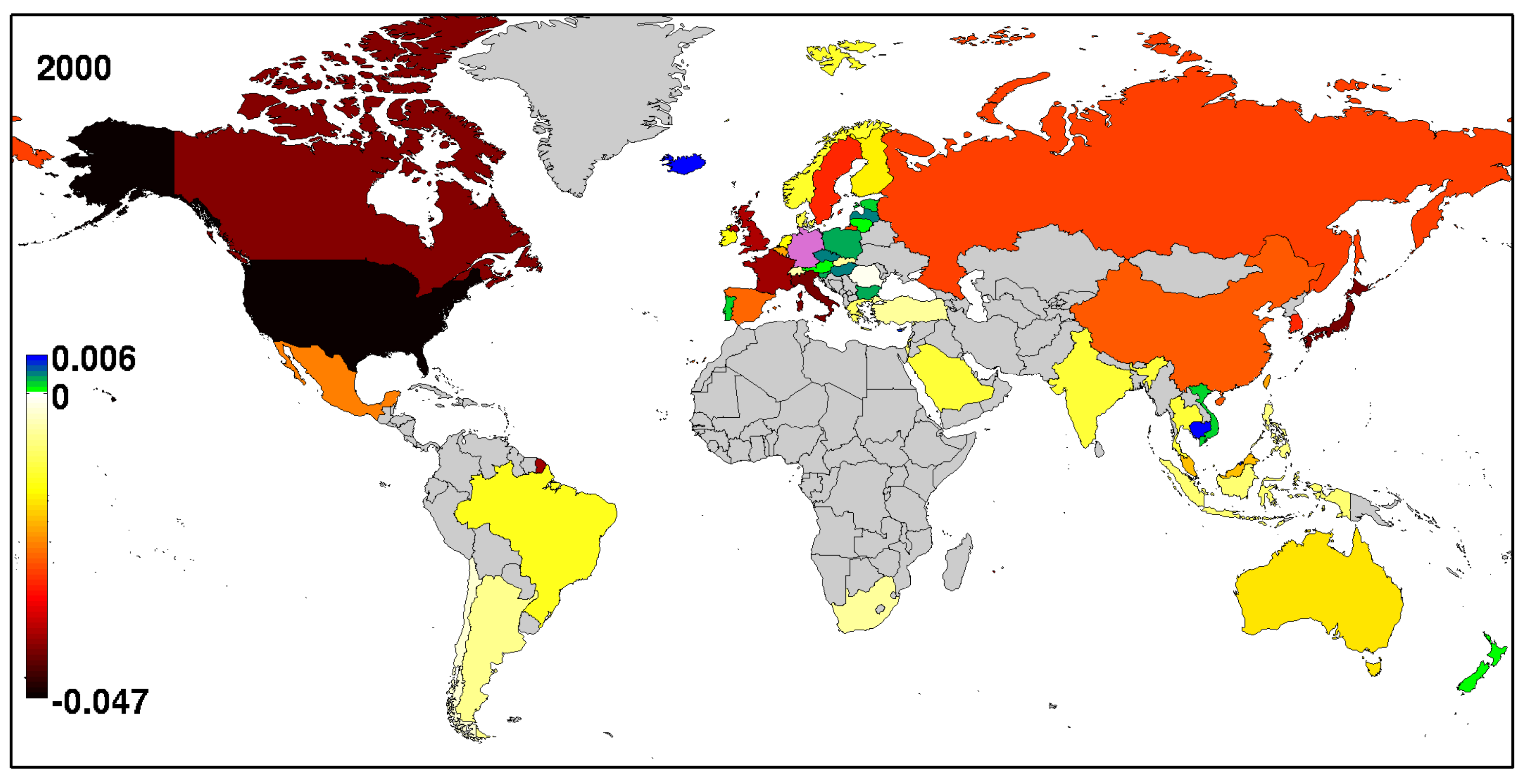} \\
\includegraphics[width=0.9\columnwidth,clip=true,trim=0 0 0 0cm]{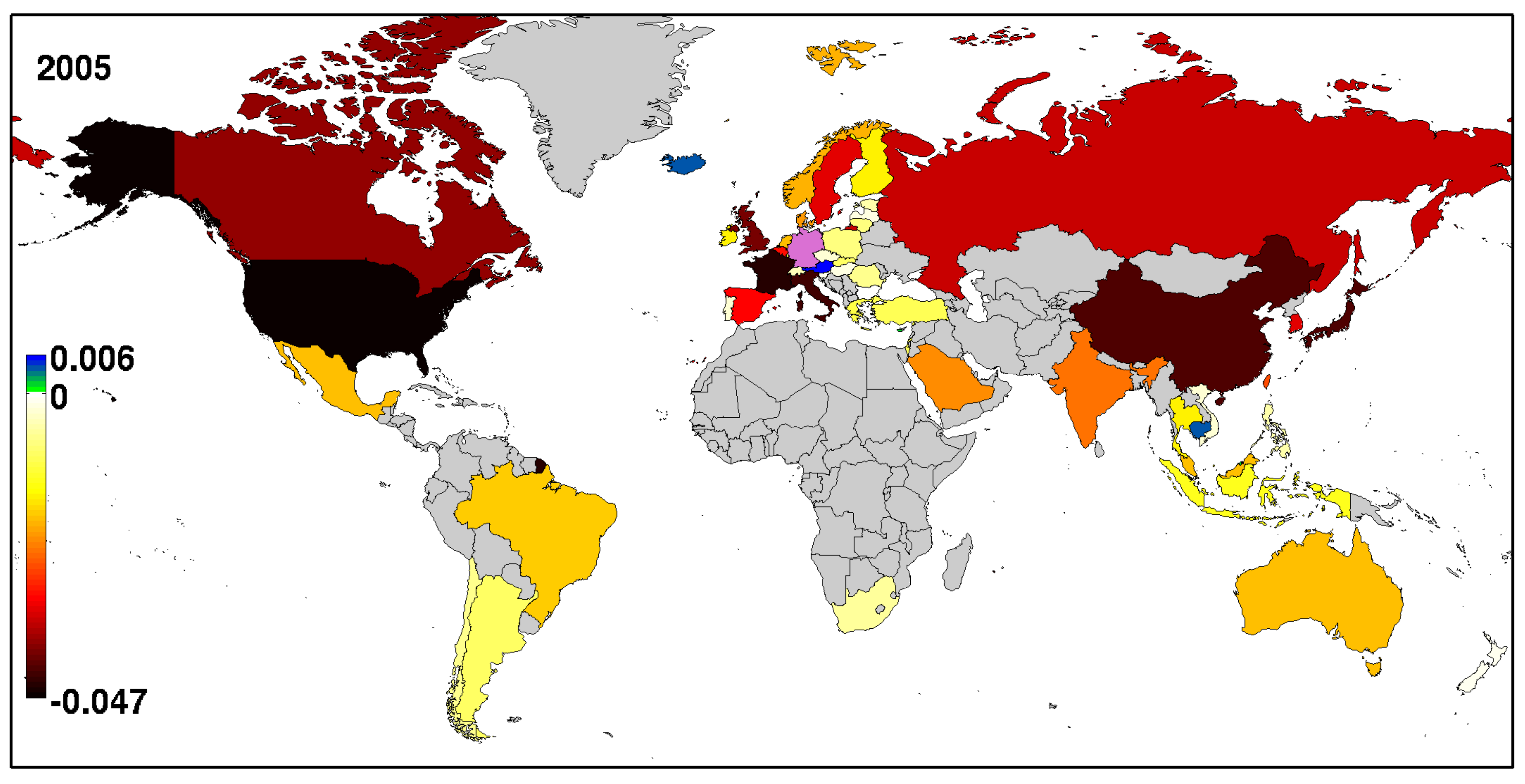} \\
\includegraphics[width=0.9\columnwidth,clip=true,trim=0 0 0 0cm]{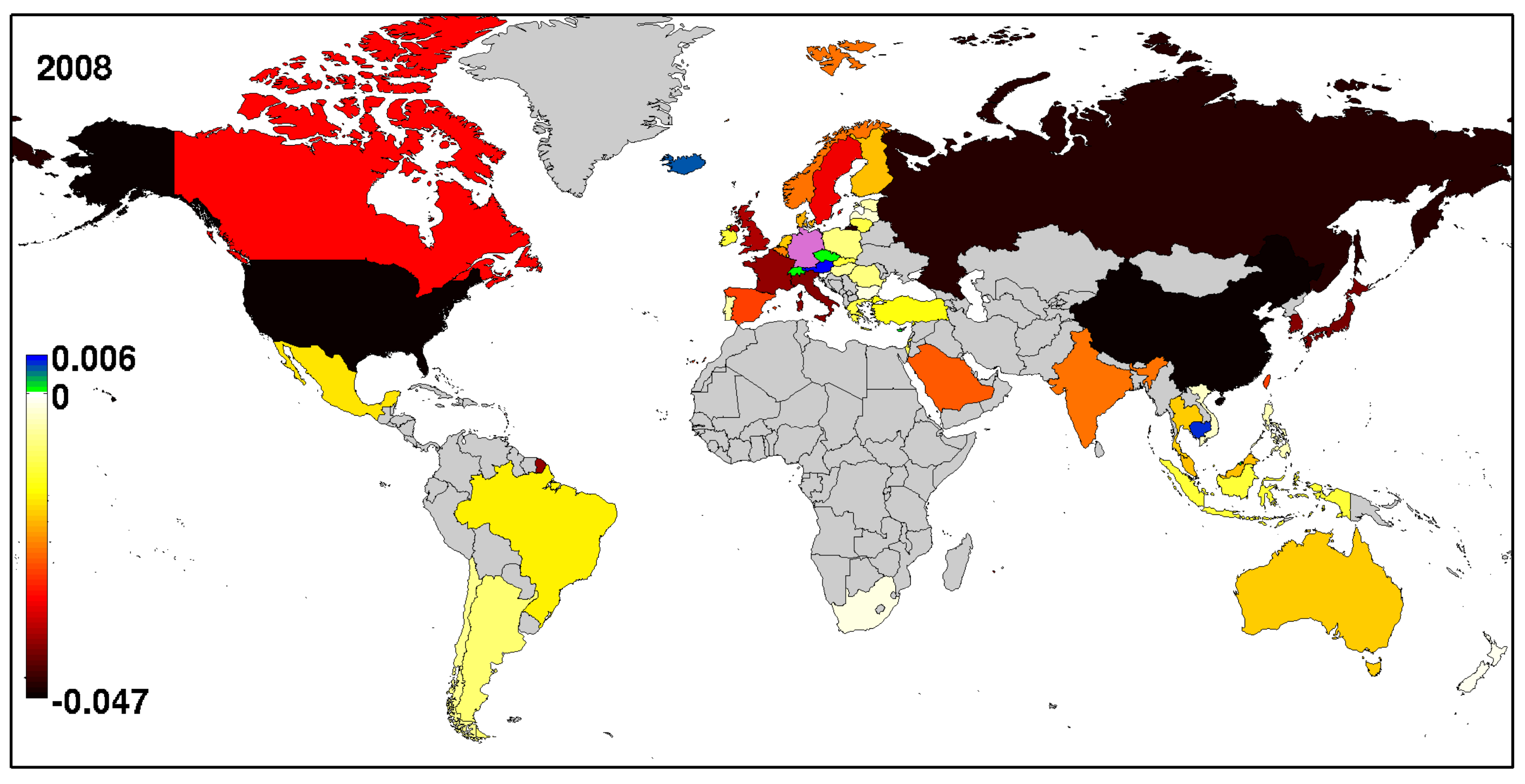} \\
\includegraphics[width=0.9\columnwidth,clip=true,trim=0 0 0 0cm]{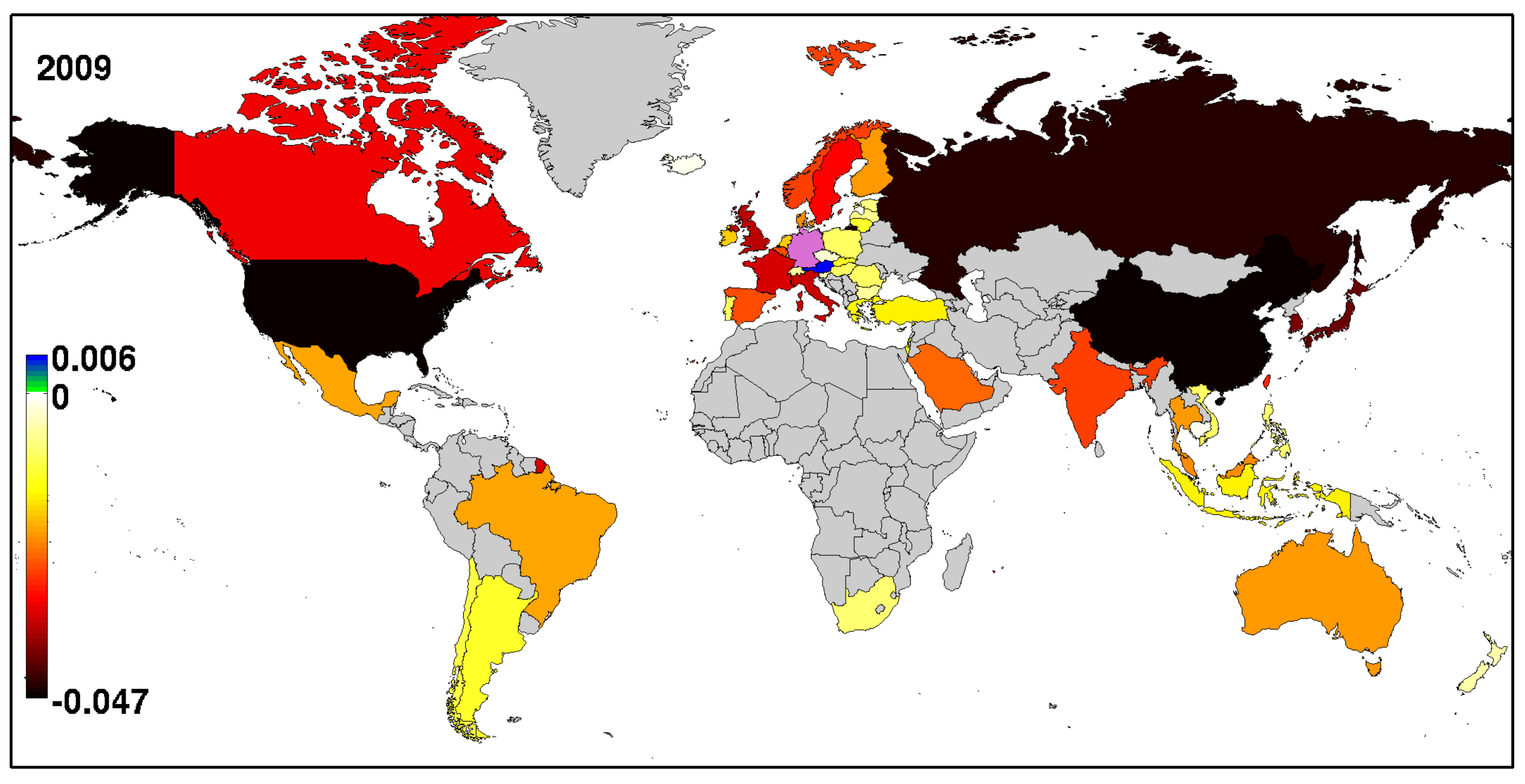}
\vglue -0.1cm
\caption {
Time evolution of the derivative $dB_c/d\sigma_{c'}$ 
over the labor cost $c'=11$ of Germany for years 1995, 2000, 2005, 2008, 2009. 
For these years the special values are 
respectively: $dB_{58}/d\sigma_{11}=-0.0402$, $-0.0307$, $-0.0351$, 
$-0.0367$, $-0.0388$ for ROW group (gray);
$dB_{11}/d\sigma_{11}=0.33$, $0.3274$, $0.3290$, $0.3248$, $0.3760$ for Germany (magenta). 
Names of the countries can be found in Table \ref{tab1} and
in the world map of countries \cite{worldmap}.}
\label{fig8}
\end{center}
\end{figure}

\begin{figure}[!ht] 
\begin{center} 
\includegraphics[width=0.9\columnwidth,clip=true,trim=0 0 0 0cm]{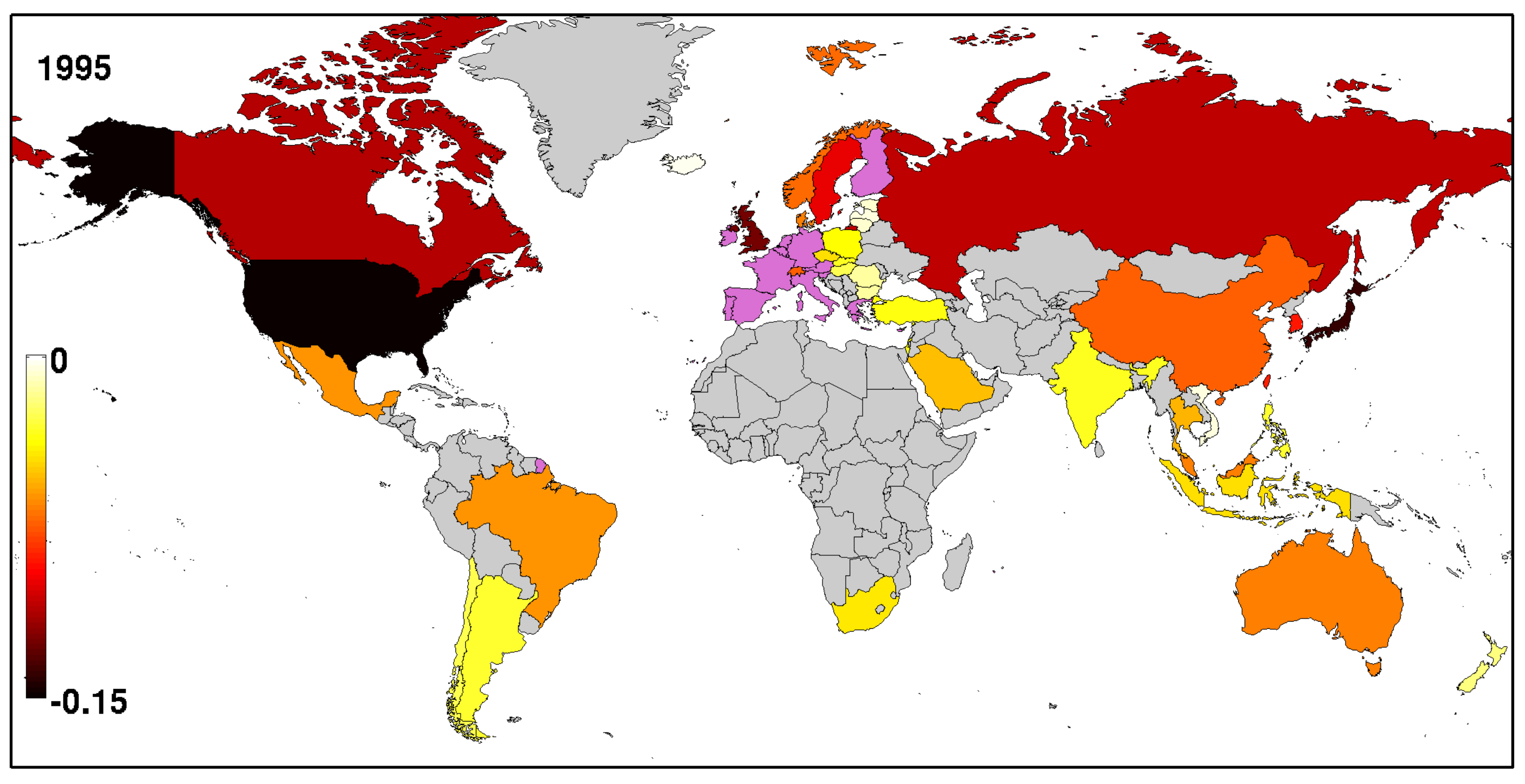} \\
\includegraphics[width=0.9\columnwidth,clip=true,trim=0 0 0 0cm]{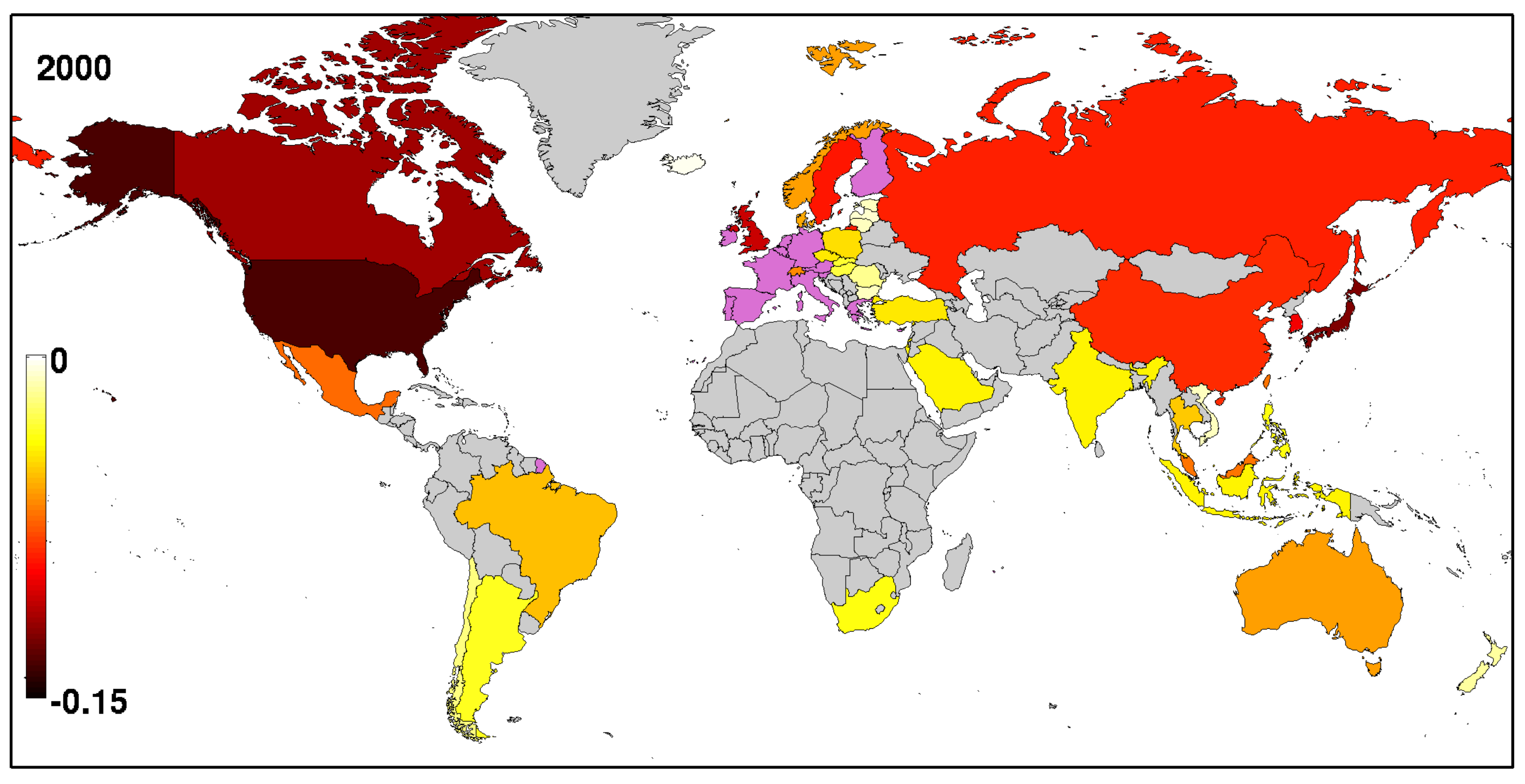} \\
\includegraphics[width=0.9\columnwidth,clip=true,trim=0 0 0 0cm]{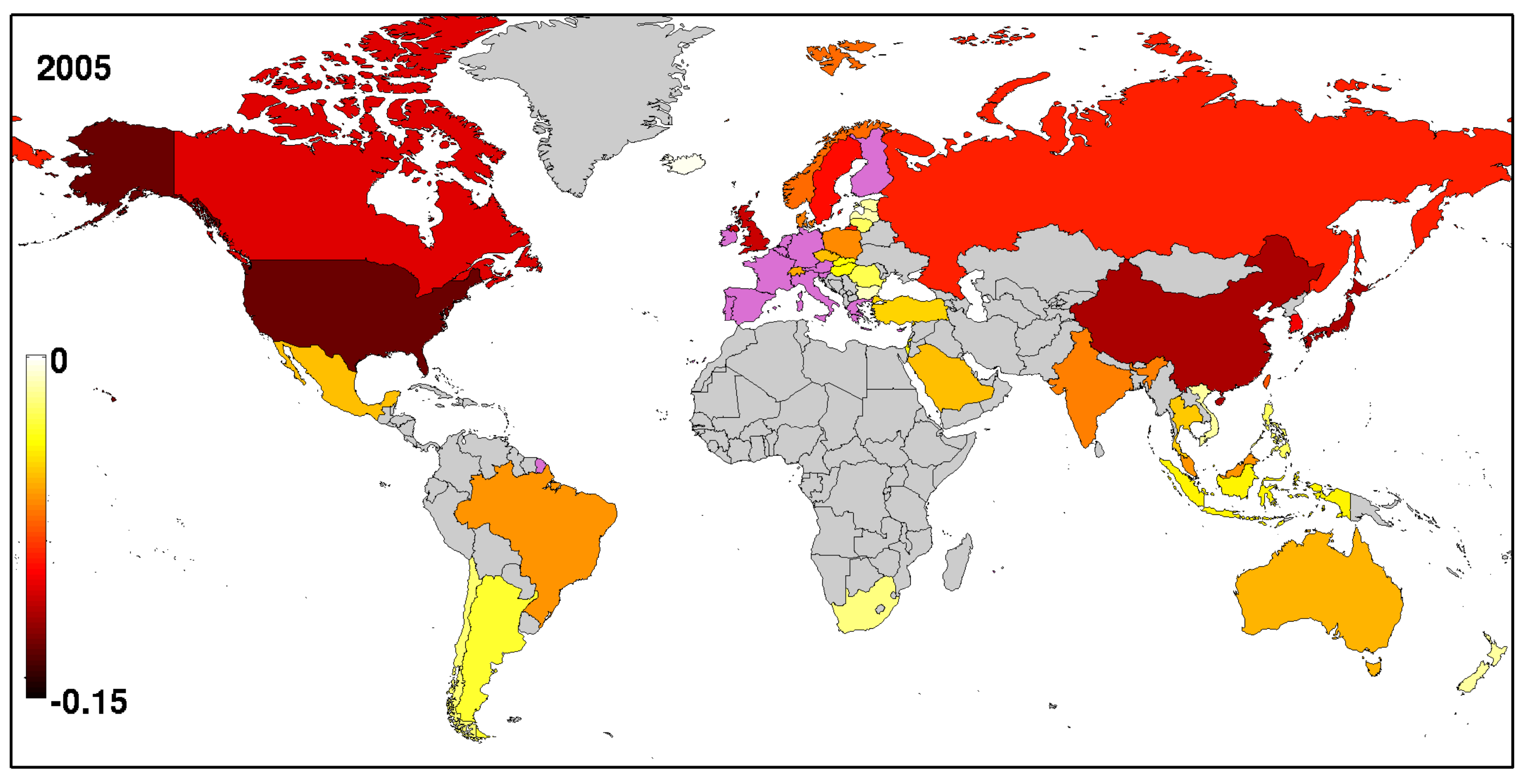} \\
\includegraphics[width=0.9\columnwidth,clip=true,trim=0 0 0 0cm]{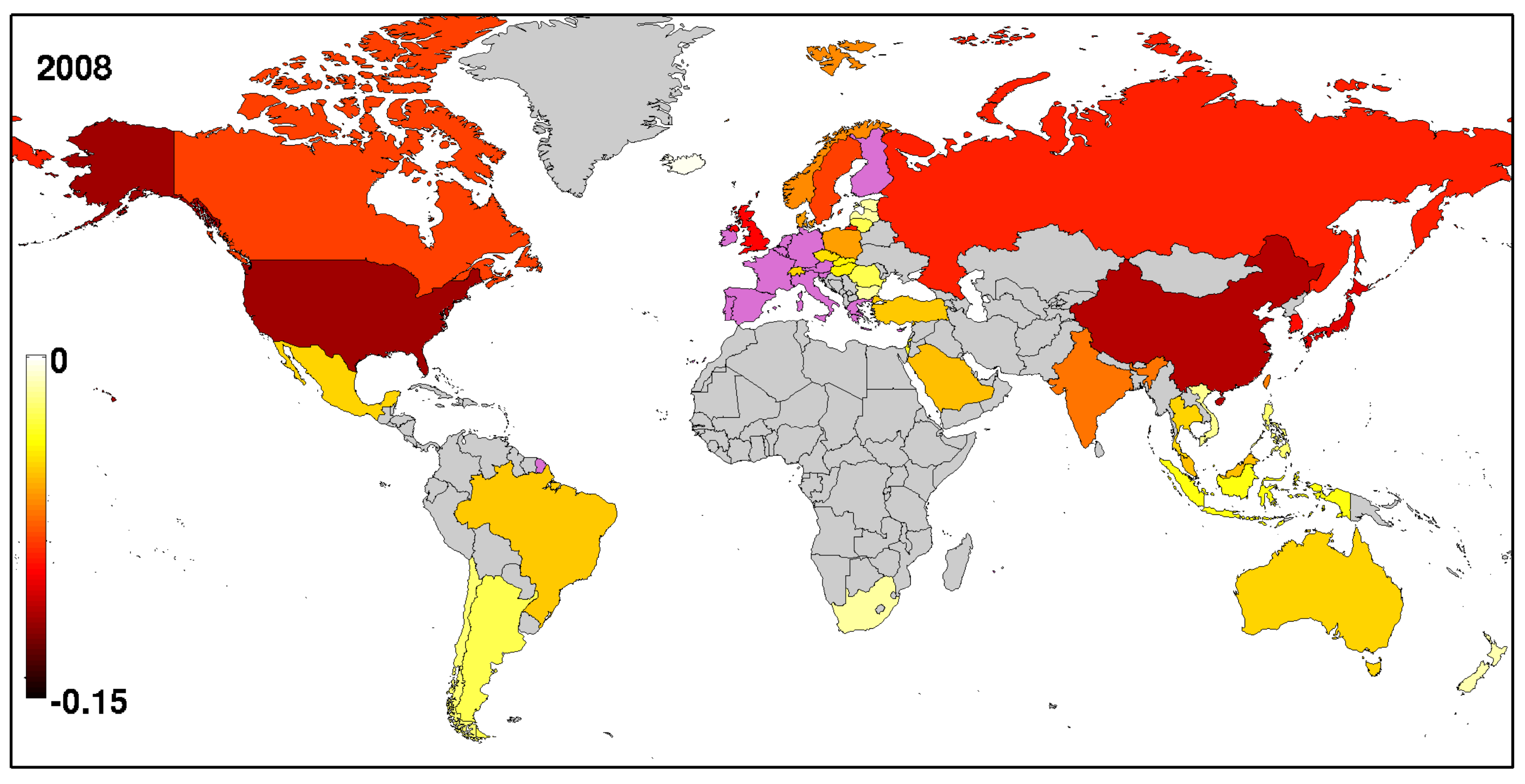} \\
\includegraphics[width=0.9\columnwidth,clip=true,trim=0 0 0 0cm]{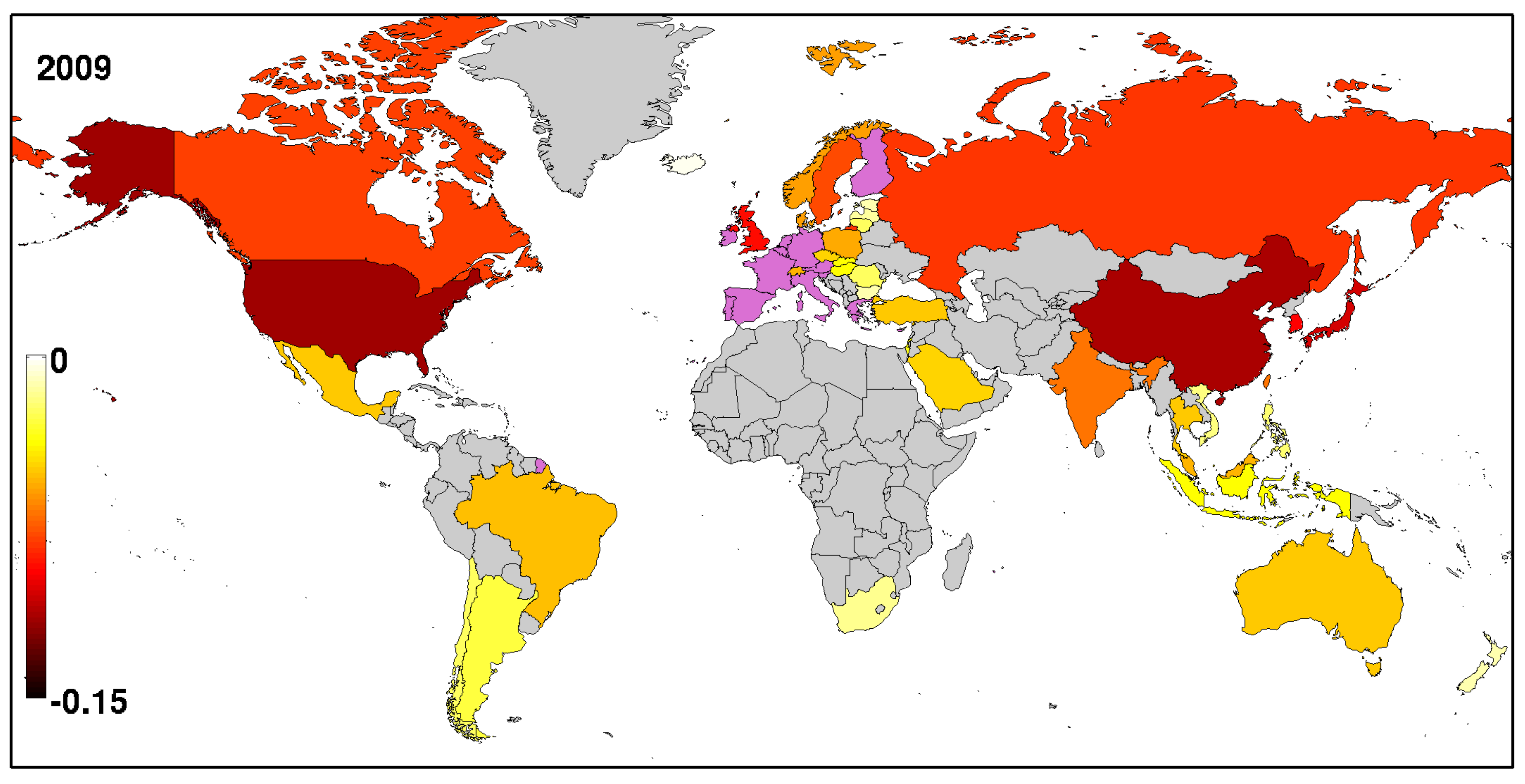}
\vglue -0.1cm
\caption {Time evolution of the derivative $dB_c/d\sigma_{c'}$ 
over the labor cost of Eurozone monetary union 
(state in 2008 composed of 15 countries) 
for years 1995, 2000, 2005, 2008, 2009.  
For these years the special values are 
respectively: $dB_{58}/d\sigma_{ez}=-0.1259$, $-0.0957$,
$-0.0992$, $-0.0908$, $-0.0921$ for ROW group (gray);
and $dB_{ez}/d\sigma_{ez}=1.8422$,
$1.9235$, $1.9394$, $1.9418$, $1.9508$ for Eurozone (magenta).
Eurozone is composed from
15 countries of its state in 2008:
Austria ($c=2$), Belgium (3), Finland (9), France (10),
Germany (11), Greece (12), Ireland (15), Italy (17),
Luxembourg (20), Netherlands (22), Portugal (26), 
Slovenia (28), Spain (29), Malta (56).
Names of the countries can be found in Table \ref{tab1} and
in the world map of countries \cite{worldmap}.
}
\label{fig9}
\end{center}
\end{figure}

\clearpage

\subsection{World map of sensitivity to labor cost}

The new element of the WNEA, compared to the multiproduct WTN,
is existence  of transfers between sectors of the same economy. This allows us to
consider the sensitivity not only to sectoral prices but also 
 the sensitivity to labor cost in a given country $c$
(e.g. price shock affecting all industries in the same country).
This can be taken into account by the introduction of
the dimensionless labor cost change in a given country $c$
by replacing the related monetary flows from coefficient $1$
to $1+\sigma_c$ in $M_{c c', s s;}$ (\ref{eq1}) for a selected country $c$.

Using the established structure of WNEA we 
can study the sensitivity of
country balance $d B_c/d \sigma_c'$ to the labor cost in different countries.
At the difference of sectoral shocks on one product,
here the price shock affects all industries in a country.
As before, the change in price has to be small enough for the
resulting simulation to remain in a neighbourhood
of the original data. In fact we compute numerically the derivate
 $d B_c/d \sigma_c'$ corresponding only to small
values of  $ \sigma_c'$.
Indeed, larger shocks would trigger
a series of substitution effects diverting
trade to other partners. The modelling in the case
of large shock variations is a very difficult task
(see discussion and analysis of such situations at \cite{escaith2}).

The derivative $d B_c/d \sigma_c'$ is computed numerically as described above.
The world sensitivity to the labor cost of China is shown 
for year 2009 in Fig.~\ref{fig7}.
Of course, the largest derivative is found for China itself
($d B_c/d \sigma_c$ at $c=37$ from Table~\ref{tab1}).
The effect on other countries is given by non-diagonal derivatives
at $c \neq c'=37$. From the CheiRank-PageRank balance we find that the most
strong negative effect (minimal negative $d B_c/d \sigma_{c'}$)
is obtained for USA, Germany, UK; a positive derivative is visible  only
for Chinese Taipei ($s=38$) and S.Korea ($s=19$).
For the Export-Import balance the results are rather different: at first all derivatives
at $c \neq c'$ are negative; among the most negative values are such countries as
Hong Kong (most negative with dark red color but hardly visible due to its small size),
Chinese Taipei,  S.Korea, Vietnam. 
This sensitivity map for year 2008 is given in \cite{wnea},
it has rather similar features.
Thus the Google matrix approach
brings a new perspective for analysis of complex of economic relations
between countries and sectors.


Another results for the time evolution of effects of labor cost in Germany 
are shown in Fig.~\ref{fig8} for all available years from 1995 to 2009.
Here we present results only for
CheiRank-PageRank balance since the results of Export-Import balance,
presented for year 2008 in \cite{wnea} do not capture 
efficiently the multiple link relations.
This time evolution demonstrates a spectacular
increase of German influence from 1995 to 2009.
Indeed, in 1995 the most strong negative sensitivity to German labor cost
is visible mainly for USA but with time the influence of
German economic activity extends to Russia and China
capturing the large fraction of the whole world.
Inside EU this influence is maximally negative in 2005
but it is slightly reduced in 2008 and 2009.

For comparison we present the same time evolution of
the sensitivity to the labor cost in
Eurozone in Fig.~\ref{fig9}.
Here, Eurozone is composed by 15 countries
present in the Euro monetary union in 2008
and the labor variation $\sigma_{c'}$ is taken to be the
same in all these countries 
when computing the derivative $d B_c/d \sigma_{c'}$
for all available years. It is striking to see that
the evolution of Eurozone influence
is opposite to Germany. Thus in 1995 we see 
a strong influence on USA which however
decreases significantly in 2008 and 2009.
The Eurozone influence on Russia (even if not so strong as for USA)
also decreases with time. There is only a certain
influence increase on China which however remains steady 
for years 2005 - 2009.


\section{Discussion}

In this work we have developed the Google matrix analysis of the 
world network of economic activities from the OECD-WTO TiVA database.
The PageRank and CheiRank probabilities allowed to obtain ranking of world countries
independently of their richness being mainly determined by the efficiency of their
economic relations.
The developed approach demonstrated the asymmetry in the economic activity sectors
some of which are export oriented and others are import oriented.

The CheiRank-PageRank balance $B_c$ allows to determine
economically rising countries with robust network of economic relations. 
The sensitivity of this $B_c$ to price variations and labor cost in various countries
determines the hidden relations between world economies being not visible via 
usual Export-Import exchange analysis. 

Our Google matrix analysis highlights the striking increase of
the influence of German economic activity on the economy of world countries
during the period 1995 to 2009. At the same time the influence
of Eurozone decreases significantly.

The knowledge of network connections in WNEA allows to investigate contagion
propagation over the whole world. Indeed, a significant
increase of petroleum prices can produce a shock wave which will propagate
over the most sensitive links highlighted in our studies.
We note that we consider the case of price contagion effect.
We plan to develop and study the models of such shock  contagion
propagation in the future works.

The comparison  with the multiproduct world trade network from UN COMTRADE
shows certain similarities between the two networks of WNEA and WTN.
At the same time the WNEA data provides new elements
for interactions of activity sectors while there are no direct 
interactions of products in COMTRADE database. From this viewpoint the OECD-WTO data
captures the economic reality on a deeper level. But at the same time the OECD-WTO
network is less developed compared to COMTRADE (less countries, years, sectors).
Thus it is highly desirable to extend the OECD-WTO database.

We think that the Google matrix analysis developed here and in \cite{wtngoogle,wtnproducts}
captures better the new reality of multifunctional directed tensor interactions
and that the universal features of this approach can be 
also extended to multifunctional financial network flows
which now attract an active interest of researchers \cite{craig,garratt}.
Unfortunately, the data on financial flows have much less 
accessibility compared to the networks discussed here.

\section{Acknowledgments}
We thank the representatives of OECD \cite{oecd2014}
and WTO \cite{wto2014} for providing us with 
the friendly access to the data sets investigated in this work.
One of us (VK) thanks 
the  Economic Research and Statistics Division, WTO Gen\`eve
for hospitality during his internship there.
We thank L.Ermann for useful discussions
and advices on preparation of figures.

\onecolumn

\clearpage

\begin{table}
\resizebox{\columnwidth}{!}{
\begin{tabular}{|c|c|c|c||c|c|c|c|} 
\hline 
 & country name & country code & country flag & & country name & country code & country flag \\ 
\hline 
\hline 
1 & Australia & AUS & \includegraphics[scale=0.4]{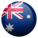} & 30 & Sweden & SWE & \includegraphics[scale=0.4]{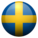}  \\ 
2 & Austria & AUT & \includegraphics[scale=0.4]{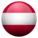} & 31 & Switzerland & CHE & \includegraphics[scale=0.4]{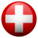}  \\ 
3 & Belgium & BEL & \includegraphics[scale=0.4]{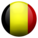} & 32 & Turkey & TUR & \includegraphics[scale=0.4]{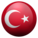}  \\ 
4 & Canada & CAN & \includegraphics[scale=0.4]{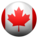} & 33 & United Kingdom & GBR & \includegraphics[scale=0.4]{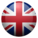}  \\ 
5 & Chile & CHL & \includegraphics[scale=0.4]{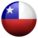} & 34 & United States & USA & \includegraphics[scale=0.4]{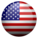}  \\ 
6 & Czech Republic & CZE & \includegraphics[scale=0.4]{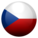} & 35 & Argentina & ARG & \includegraphics[scale=0.4]{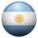}  \\ 
7 & Denmark & DNK & \includegraphics[scale=0.4]{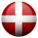} & 36 & Brazil & BRA & \includegraphics[scale=0.4]{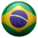}  \\ 
8 & Estonia & EST & \includegraphics[scale=0.4]{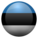} & 37 & China & CHN & \includegraphics[scale=0.4]{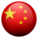}  \\ 
9 & Finland & FIN & \includegraphics[scale=0.4]{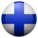} & 38 & Chinese Taipei & TWN & \includegraphics[scale=0.4]{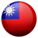}  \\ 
10 & France & FRA & \includegraphics[scale=0.4]{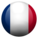} & 39 & India & IND & \includegraphics[scale=0.4]{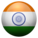}  \\ 
11 & Germany & DEU & \includegraphics[scale=0.4]{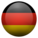} & 40 & Indonesia & IDN & \includegraphics[scale=0.4]{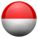}  \\ 
12 & Greece & GRC & \includegraphics[scale=0.4]{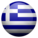} & 41 & Russia & RUS & \includegraphics[scale=0.4]{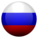}  \\ 
13 & Hungary & HUN & \includegraphics[scale=0.4]{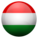} & 42 & Singapore & SGP & \includegraphics[scale=0.4]{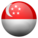}  \\ 
14 & Iceland & ISL & \includegraphics[scale=0.4]{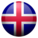} & 43 & South Africa & ZAF & \includegraphics[scale=0.4]{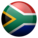}  \\ 
15 & Ireland & IRL & \includegraphics[scale=0.4]{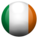} & 44 & Hong Kong & HKG & \includegraphics[scale=0.4]{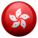}  \\ 
16 & Israel & ISR & \includegraphics[scale=0.4]{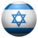} & 45 & Malaysia & MYS & \includegraphics[scale=0.4]{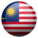}  \\ 
17 & Italy & ITA & \includegraphics[scale=0.4]{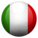} & 46 & Phillippines & PHL & \includegraphics[scale=0.4]{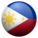}  \\ 
18 & Japan & JPN & \includegraphics[scale=0.4]{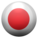} & 47 & Thailand & THA & \includegraphics[scale=0.4]{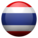}  \\ 
19 & Korea & KOR & \includegraphics[scale=0.4]{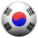} & 48 & Romania & ROU & \includegraphics[scale=0.4]{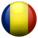}  \\ 
20 & Luxembourg & LUX & \includegraphics[scale=0.4]{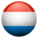} & 49 & Vietnam & VNM & \includegraphics[scale=0.4]{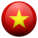}  \\ 
21 & Mexico & MEX & \includegraphics[scale=0.4]{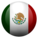} & 50 & Saudi Arabia & SAU & \includegraphics[scale=0.4]{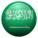}  \\ 
22 & Netherlands & NLD & \includegraphics[scale=0.4]{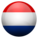} & 51 & Brunei Darussalam & BRN & \includegraphics[scale=0.4]{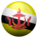}  \\ 
23 & New Zealand & NZL & \includegraphics[scale=0.4]{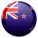} & 52 & Bulgaria & BGR & \includegraphics[scale=0.4]{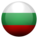}  \\ 
24 & Norway & NOR & \includegraphics[scale=0.4]{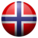} & 53 & Cyprus & CYP & \includegraphics[scale=0.4]{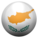}  \\ 
25 & Poland & POL & \includegraphics[scale=0.4]{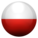} & 54 & Latvia & LVA & \includegraphics[scale=0.4]{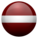}  \\ 
26 & Portugal & PRT & \includegraphics[scale=0.4]{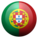} & 55 & Lithuania & LTU & \includegraphics[scale=0.4]{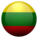}  \\ 
27 & Slovak Republic & SVK & \includegraphics[scale=0.4]{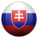} & 56 & Malta & MLT & \includegraphics[scale=0.4]{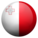}  \\ 
28 & Slovenia & SVN & \includegraphics[scale=0.4]{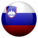} & 57 & Cambodia & KHM & \includegraphics[scale=0.4]{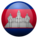}  \\ 
29 & Spain & ESP & \includegraphics[scale=0.4]{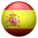} & 58 & Rest of the World & ROW & \includegraphics[scale=0.4]{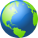}  \\ 
\hline 
\end{tabular} 
}
\caption{List of $N_c=58$ countries (with rest of the world ROW) with country name, code and flag.}
\label{tab1}
\end{table}

\clearpage

\begin{table}
\resizebox{\columnwidth}{!}{
\begin{tabular}{|c|c|l|} 
\hline 
 & OECD ICIO Category & ISIC Rev. 3 correspondence \\ 
\hline 
\hline 

1 &  C01T05 AGR	& \shortstack[l]{ 01 - Agriculture, hunting and related service activities \\
		  02 - Forestry, logging and related service activities \\
		  05 - Fishing, operation of fish hatcheries and fish farms; service activities incidental to fishing} \\ \hline
2 &  C10T14 MIN	& \shortstack[l]{ 10 - Mining of coal and lignite; extraction of peat \\
		  11 - Extraction of crude petroleum and natural gas; service activities incidental to oil and gas extraction excluding surveying \\
		  12 - Mining of uranium and thorium ores \\
		  13 - Mining of metal ores \\
		  14 - Other mining and quarrying} \\ \hline
3 &  C15T16 FOD & \shortstack[l]{ 15 - Manufacture of food products and beverages \\
		  16 - Manufacture of tobacco products} \\ \hline
4 &  C17T19 TEX	& \shortstack[l]{ 17 - Manufacture of textiles \\
		  18 - Manufacture of wearing apparel; dressing and dyeing of fur \\
		  19 - Tanning and dressing of leather; manufacture of luggage, handbags, saddlery, harness and footwear} \\ \hline
5 &  C20 WOD	& \shortstack[l]{ 20 - Manufacture of wood and of products of wood and cork, except furniture; \\
                       Manufacture of articles of straw and plaiting materials} \\ \hline
6 &  C21T22 PAP	& \shortstack[l]{ 21 - Manufacture of paper and paper products \\
		  22 - Publishing, printing and reproduction of recorded media} \\ \hline
7 &  C23 PET	& 23 - Manufacture of coke, refined petroleum products and nuclear fuel \\ \hline
8 &  C24 CHM	& 24 - Manufacture of chemicals and chemical products \\ \hline
9 &  C25 RBP	& 25 - Manufacture of rubber and plastics products \\ \hline
10 & C26 NMM	& 26 - Manufacture of other non-metallic mineral products \\ \hline
11 & C27 MET	& 27 - Manufacture of basic metals \\ \hline
12 & C28 FBM	& 28 - Manufacture of fabricated metal products, except machinery and equipment \\ \hline
13 & C29 MEQ	& 29 - Manufacture of machinery and equipment n.e.c. \\ \hline
14 & C30 ITQ	& 30 - Manufacture of office, accounting and computing machinery \\ \hline
15 & C31 ELQ	& 31 - Manufacture of electrical machinery and apparatus n.e.c. \\ \hline
16 & C32 CMQ	& 32 - Manufacture of radio, television and communication equipment and apparatus \\ \hline
17 & C33 SCQ	& 33 - Manufacture of medical, precision and optical instruments, watches and clocks \\ \hline
18 & C34 MTR	& 34 - Manufacture of motor vehicles, trailers and semi-trailers \\ \hline
19 & C35 TRQ	& 35 - Manufacture of other transport equipment \\ \hline
20 & C36T37 OTM	& \shortstack[l]{ 36 - Manufacture of furniture; manufacturing n.e.c. \\
		  37 - Recycling} \\ \hline
21 & C40T41 EGW	& \shortstack[l]{ 40 - Electricity, gas, steam and hot water supply \\
		  41 - Collection, purification and distribution of water} \\ \hline
22 & C45 CON	& 45 - Construction \\ \hline
23 & C50T52 WRT	& \shortstack[l]{ 50 - Sale, maintenance and repair of motor vehicles and motorcycles; retail sale of automotive fuel \\
		  51 - Wholesale trade and commission trade, except of motor vehicles and motorcycles \\
		  52 - Retail trade, except of motor vehicles and motorcycles; repair of personal and household goods} \\ \hline
24 & C55 HTR	& 55 - Hotels and restaurants \\ \hline
25 & C60T63 TRN	& \shortstack[l]{ 60 - Land transport; transport via pipelines \\
		  61 - Water transport \\
		  62 - Air transport \\
		  63 - Supporting and auxiliary transport activities; activities of travel agencies} \\ \hline
26 & C64 PTL	& 64 - Post and telecommunications \\ \hline
27 & C65T67 FIN	& \shortstack[l]{ 65 - Financial intermediation, except insurance and pension funding \\
		  66 - Insurance and pension funding, except compulsory social security \\
		  67 - Activities auxiliary to financial intermediation} \\ \hline
28 & C70 REA	& 70 - Real estate activities \\ \hline
29 & C71 RMQ	& 71 - Renting of machinery and equipment without operator and of personal and household goods \\ \hline
30 & C72 ITS	& 72 - Computer and related activities \\ \hline
31 & C73 RDS	& 73 - Research and development \\ \hline
32 & C74 BZS	& 74 - Other business activities \\ \hline
33 & C75 GOV	& 75 - Public administration and defense; compulsory social security \\ \hline
34 & C80 EDU	& 80 - Education \\ \hline
35 & C85 HTH	& 85 - Health and social work \\ \hline
36 & C90T93 OTS	& \shortstack[l]{ 90 - Sewage and refuse disposal, sanitation and similar activities \\
		  91 - Activities of membership organizations n.e.c. \\
		  92 - Recreational, cultural and sporting activities \\
		  93 - Other service activities} \\ \hline
37 & C95 PVH	& 95 - Private households with employed persons \\ \hline
\end{tabular}
} 
\caption{List of sectors considered by Input/Output matrices from OECD database, their correspondence to the ISIC classification is also given.}
\label{tab2}
\end{table}

\end{document}